\documentclass[conference]{IEEEtran} 
\usepackage{url}
\usepackage{amsmath}
\usepackage{amsfonts}
\usepackage{amssymb}
\usepackage{subcaption}
\usepackage{pifont}
\usepackage{balance}

\usepackage{array}

\usepackage{enumitem}
\setlist[itemize]{noitemsep, topsep=0.1cm}
\setlist[enumerate]{noitemsep, topsep=0.1cm}
\usepackage{caption}

\usepackage{color}
\usepackage{soul}
\usepackage{multirow}
\usepackage{comment}

\newcommand{\tabred}{-4pt}

\newcommand{\topnumbers}{} 

\definecolor{lightcyan}{RGB}{210, 210, 250}

\newcommand{\hilight}[1]{#1}

\usepackage{graphicx}
\graphicspath{{./}}
\DeclareGraphicsExtensions{.png,.jpeg,.pdf}

\begin{document}\sloppy
%

\title{CAPTCHaStar! A Novel CAPTCHA \\ Based on Interactive Shape Discovery%
}
%
%
%
%
%

\author{
\IEEEauthorblockN{ Mauro Conti, Claudio Guarisco and Riccardo Spolaor}
\IEEEauthorblockA{University of Padua, Padua, Italy\\ \{conti, rspolaor\}@math.unipd.it, cguarisc@gmail.com}
}

\maketitle

\begin{abstract}
Over the last 
years, 
most websites on which users can register (e.g., email providers and social networks)
adopted CAPTCHAs (Completely Automated Public Turing test to tell Computers and Humans Apart) 
as a countermeasure against automated attacks. 
The battle of wits between designers and attackers of CAPTCHAs led to current ones being annoying and hard to solve for users, while still being vulnerable to automated attacks.

In this paper, we propose CAPTCHaStar, a new image-based CAPTCHA that relies on user interaction.
This novel CAPTCHA leverages 
the innate human ability to recognize shapes in a confused environment.
We assess the effectiveness of our proposal 
for the two key aspects for CAPTCHAs, i.e., usability, and resiliency to automated attacks.
In particular, we evaluated the usability, carrying out a thorough user study, and we tested the resiliency of our proposal against several types of automated attacks: traditional ones; designed ad-hoc for our proposal; and based on machine learning.
Compared to the state of the art, our proposal is more user friendly (e.g.,
only some 35\% of the users prefer current solutions, such as text-based CAPTCHAs)
and more resilient to automated attacks.
\end{abstract}

\section{Introduction}
Many public services on the Internet are 
subject to automated attacks, i.e., 
an automated program can exploit a vulnerable on-line service, pretending to be a legitimate user.
As an example, an attacker may create multiple accounts on an e-mail provider and use them to send spam messages. 
In the last years, an 
increasing number of websites adopted countermeasures against these malicious attacks.
The most common method consists in allowing access to a service only to users able to solve a CAPTCHA 
(Completely Automated Public Turing Test to Tell Computers and Humans Apart). 
The main purpose of a CAPTCHA is to distinguish a human user from a 
software robot (from now on also referred as ``bot'') that runs automated tasks.
In order to do that, researchers leverage the existing gap between human abilities 
and the current state of the art of 
software, including also Artificial Intelligence techniques~\cite{lopresti2005leveraging}. 
A CAPTCHA is a program that generates a test, which has the property to be easily solvable by humans, but hardly solvable by a bot~\cite{von2004telling}
(if not employing a significant amount of resources and time).
As an example, a bot cannot easily understand the meaning of a sentence (or a picture), while humans can carry out this task with negligible effort. 
  
The design of a good CAPTCHA is not a trivial task.
Indeed, both usability to legitimate users and resiliency against automated attacks must be simultaneously satisfied.
Attackers of CAPTCHA usually improve automated attacks over time.
For this reason, designers use to improve their CAPTCHAs in order to reduce the success rate of novel attacks.
Unfortunately, these improvements usually cause a dramatic decrease in usability~\cite{bursztein2010good}. 
Researchers put a significant effort in understanding the trade-off between usability and resiliency to attacks~\cite{ben2009experimental}. 
\label{usabilitymetrics}
Also, in order to measure the effective usability of a CAPTCHA, Yan et al.~\cite{yan2008usability} presented a set of metrics that we also consider in this paper: \textit{success rate}, \textit{completion time} and \textit{ease of understanding}.

\noindent
\textit{Contribution}. 
The contribution of this paper is as follow:
\begin{itemize}
   \item We present CAPTCHaStar\footnote{A demo is available at \url{http://captchastar.math.unipd.it/demo.php}.}, a novel CAPTCHA based on shape recognition and user interaction. 
  CAPTCHaStar prompts the user with some ``stars'' inside a square. 
  The position of these stars changes according to the position of the cursor. 
  The user must move the cursor, until the stars aggregate in a recognizable shape. 
  Our CAPTCHA leverages the innate human ability to recognize a shape in a confused environment. 
  Indeed, a machine cannot easily emulate this ability~\cite{hinton2007recognize}.
  This makes CAPTCHaStar easy solvable by humans while remaining difficult for bots. 
%
  \item 
We assessed the usability of our proposal via a user study, considering an extensive set of parameters.
The results show that CAPTCHaStar users achieve a success rate higher than $90\%$ 
  for the best combination of parameter values. 
  Furthermore, humans can solve our CAPTCHA in less than $30$ seconds (on average).
%
  \item
  We assess the security of our proposal.
  In particular, we first studied the resiliency of our CAPTCHaStar against traditional attacks (such as exhaustion and leak of the database). 
  Then, we present some possible ad-hoc attack strategies and discuss their effectiveness against our proposal.
  Finally, we also assessed the resiliency of CAPTCHaStar against attacks based on machine learning.
  In all these studies, our solution showed promising results, comparable or even better than state of the art solutions.
  \item We compare the features of CAPTCHaStar with other existing CAPTCHAs.
  In particular, we compare our proposal against some
  of the most famous image-based designs in the literature.
  For each of these designs, we discuss the protection that it offers against various attack strategies. 
  The results of our comparison underline that our design improves the state of the art.
\end{itemize}
Our work suggests that CAPTCHaStar is promising for a practical wide adoption (particularly for mobile devices, where the use of keyboard is more difficult and error-prone~\cite{reynaga2013usability}), as well as motivate further research along the same direction.

\noindent
\textit{Organization}. The rest of this paper is organized as follows.
In Section~\ref{relatedwork}, we report an overview of the current state of the art.
In Section~\ref{ourproposed}, we describe in details CAPTCHaStar, our novel CAPTCHA.
In Section~\ref{evaluateusability}, we evaluate its usability features, 
while in Section~\ref{evaluateresiliency}, we assess its resiliency to automated attacks.
In Section~\ref{discussion}, we compare CAPTCHaStar with other image-based CAPTCHAs in the literature, and we discuss limitatons and possible future work.
Finally, in Section~\ref{conclusions}, we draw some conclusions summarizing the contributions of our research.

\section{Related work}
\label{relatedwork}
In this section, we discuss the main techniques in the literature to design CAPTCHAs, along with their pros and cons.
This section is not intended to be a comprehensive review of the whole literature.
Interested readers can refer to the work in~\cite{shiralidistinguishing} for an extensive survey of the state of the art.
Henceforth, we refer to a single instance of a CAPTCHA test prompted to a user with the term \textit{challenge}.
In the following sections, we divide CAPTCHAs in two main categories, according to the skill required to solve them: 
\textit{text-based} (Section~\ref{tbcaptcha}), when they require text recognition, and \textit{image-based} (Section~\ref{imagebasedcompetitors}), when they 
challenge the user to recognize images. 
For each category, we briefly describe 
their usability, traditional attack strategies, and possible countermeasures. 
Recently, Google proposed noCaptcha, a system that uses an ``advanced risk analysis back-end that considers the engagement of the user'' and prompts the user with either a text-based or an image-based challenge~\cite{googlenocaptcha}. Unfortunately, there is not yet much technical information available (as well as research papers) to understand how exactly it works, nor to run a proper comparison. 
As far as we know, the actual CAPTCHA prompted to the user seems independent from the actual ``risk assessment'', i.e., even CAPTCHaStar might be used!
\subsection{Text-based CAPTCHAs}
\label{tbcaptcha}
A text-based CAPTCHA presents an obfuscated word in the form of an image, and asks the user to read and rewrite it, usually in a text box.
Baird et al.~\cite{baird2003pessimalprint} proposed the first text-based CAPTCHA in 2002.
After this first proposal, several other researchers worked on this kind of design. 
Several researchers focused on improving the resiliency against automated attacks~\cite{baird2005scattertype,ferzli2006captcha,ince2008designing}. 
Currently, text-based CAPTCHAs are the most widely used~\cite{bursztein2011text}.
In the following, we report two examples of text-based CAPTCHAs that, as for CAPTCHaStar, do not require a keyboard to submit the answer: 
iCaptcha~\cite{truong2011iCAPTCHA} and DDIM-CAPTCHA~\cite{ye2013ddim}.

The authors of iCaptcha~\cite{truong2011iCAPTCHA} focused their efforts on the prevention of relay attacks~\cite{motoyama2010re}:
i.e., when a bot uses an external paid human to solve a CAPTCHA. 
This text-based CAPTCHA measures and analyzes the interactions a user performs while solving the challenge. 
iCaptcha prompts the user with a classical obfuscated word. 
For each word, there is a sequence of obfuscated letters that the user has to use to compose her answer.
iCaptcha verification operates on two fronts. 
First, the correctness of the answer discriminates a real human from a machine. 
Second, the interleaving time after the tap 
of each button discriminates a legitimate user from an external paid human.
However, we consider this type of discrimination weak, because the latency of the network connection can heavily affect the measurement of the interleaving times.
Moreover, iCaptcha presents the user a small set of candidate characters (i.e., the set of buttons) that  composes the solution of the challenge.
Unfortunately, while this feature  improves usability, it also increases the success rate of attacks that leverage OCR (Optical Character Recognition) software. 

The CAPTCHA design proposed by Ye et al.~\cite{ye2013ddim} (named DDIM-CAPTCHA) also leverages obfuscated words like traditional text-based schemes.
The main difference with the previous design is in 
the way the user can solve a challenge: instead of asking the user to type the answer, this CAPTCHA asks to drag the correct letters from a pool into an ``answer box''.
Letters in the pool overlap to each other, so the user has to constantly interact with the pool to pick the letter he wants. 

\paragraph{Usability features} The first implementations 
of text-based CAPTCHAs had 
a very short completion time and high success rate for legitimate users. 
Unfortunately, the introduction of countermeasures to new automated attacks have dramatically lowered these usability features, highlighting the need for new designs~\cite{fidas2011necessity}. 
The instructions to solve text-based CAPTCHAs are really easy to understand.
Indeed, they usually do not need any a-priori 
knowledge from the users, except for the ability to read.
Users need to type the answer using a keyboard, except for particular designs (e.g., iCaptcha~\cite{truong2011iCAPTCHA}). 
Unfortunately, inputting the answer with a keyboard undermine the usability of a CAPTCHA on smartphone or tablet.
Indeed, in such devices, a single-handed touch-based interaction style is dominant~\cite{posladubiquitous}. 

\paragraph{Attacks and countermeasures}
The most common way to automatically solve text-based CAPTCHAs is to use an OCR (Optical Character Recognition) software. 
In the past few years, CAPTCHAs designers and attackers took part in a battle of wits. 
This battle led to an improvement of OCR software, 
hence making OCR a very effective threat~\cite{chellapilla2005computers} to text-based CAPTCHAs.
%
%
Another effective approach to solve CAPTCHA is the so-called \textit{relay attack}: some companies sells real-time human labor to solve CAPTCHAs~\cite{motoyama2010re}.
This approach has a really high success rate and it costs only one U.S. dollar per thousand CAPTCHAs~\cite{bursztein2010good}. 

Looking at the literature, the attack strategies against text-based CAPTCHAs can be classified as follows:
\begin{enumerate}[leftmargin=1.cm]
  \item[A01)] Forward the challenge to paid or unaware humans that solve it (i.e., relay attack).
  \item[A02)] In case the answer is a word of sense, use OCR technology combined with a dictionary. 
  \item[A03)] Use OCR software on a single character separately.
  \item[A04)] Segment the word, in order to obtain a single image for every character.
  \item[A05)] Remove smaller lines that can be added as an obstacle to the segmentation process. 
  \item[A06)] Fill hollow spaces inside each character, to improve OCR effectiveness.
  \item[A07)] Repair characters outline by fixing broken lines. This method leverages on analyzing the distance between pixels. 
\end{enumerate}
Attackers may combine two or more of these attack strategies in order to achieve a higher success rate.

CAPTCHA designers reacted to these attacks proposing several improvements to mitigate their effectiveness.
Some examples follow (between parenthesis we indicate the attack 
for which the mitigation strategy could be effective):
\begin{itemize}
  \item Add more layers of interaction between user and CAPTCHA (could be effective for threat A01 above).
\item Add more distortion to the letters, e.g., warping, scaling, rotating (against A02 and A06).
\item Use of English-like words (for the sake of usability) or totally random words (against A03).
  \item Add more pollution to the image, e.g., ticker lines over the letters, confusing background (against A04 and A05).
  \item Increment noise, e.g., degrading the quality of the resulting image (against A07).
\end{itemize}
Unfortunately, some of these mitigation strategies have been shown to be ineffective~\cite{ElAhmad:2010:RNC:1752046.1752052,bursztein2014end}.

\subsection{Image-based CAPTCHAs}
\label{imagebasedcompetitors}
\label{imagebasedthreats}
Image-based CAPTCHAs usually ask the user to recognize an image or to  interact with on-screen objects 
to find a solution.
Unlike text-based CAPTCHAs, every image-based design is substantially different from each other. 
For this reason, a user who faces a CAPTCHA design for the first time needs a little more effort to understand how it works. 
Studies suggest that image-based CAPTCHAs are more appreciated by users~\cite{gao2010novel}. 
Indeed, image-based CAPTCHAs usually have a high success rate and they are less challenging than text-based ones~\cite{nejati2014deepCAPTCHA}.
In the following, we report some examples of image-based CAPTCHA that we could group in three sub-categories: static, motion, and interactive. 

One of the representative static image-based CAPTCHAs was {Asirra}~\cite{elson2007asirra}, which was discontinued in fall 2014. 
Asirra asks the user to distinguish between cats and dogs, on twelve different photos randomly taken from an external website. 
Another static image-based CAPTCHA is Collage~\cite{shirali2008advanced}: it requests to click on a specific picture, among six pictures randomly taken.
{Deep CAPTCHA}~\cite{nejati2014deepCAPTCHA} prompts the user with six 3D models of real world objects and it asks to sort them by their size. 

Some designers focus on CAPTCHA that requires video recognition rather than static image recognition.
For example, {Motion CAPTCHA}~\cite{shirali2008motion} shows the user a randomly chosen video from a database, then it asks the user to identify the action performed by the person in the video.
Similarly, {YouTube Videos CAPTCHA}~\cite{kluever2009balancing} leverages on real video in YouTube service, and it asks the user to write three tags related to the content of the video.

Interactive CAPTCHAs mitigate the relay attack threat.
For example, {Noise CAPTCHA}~\cite{okada2012new} presents a transparent noisy image overlapped to a noisy background.
The user needs to drag this image until he can recognize a well formed text. 
{Cursor CAPTCHA}~\cite{thomas2013cursor} changes the appearance of mouse cursor into another random object.
The user needs to overlap the cursor on the identical object placed in a random generated image.
{Jigsaw CAPTCHA}~\cite{gao2010novel} reprises the classical jigsaw puzzle.
Indeed, the user needs to correctly rearrange the pieces of a jigsaw.
Finally, {PlayThru}~\cite{areyouahuman} asks the user to solve a randomly generated mini-game.
These mini-games require to drag objects on their correct spots.

\paragraph{Usability features}
Since image-based CAPTCHAs are different from each other, the usability may change depending on the considered design.
Usually, image-based CAPTCHAs do not require to type on a keyboard.
For this reason, smartphone and tablet users prefer image-based CAPTCHAs over text-based ones~\cite{reynaga2013usability}.
The instructions for each different CAPTCHA design are usually short and intuitive. 
Finally, on the server-side, resources required and setup time should be as small as possible. 
However, some image-based CAPTCHAs need many external libraries and 
may require a large amount of computational power (for example, the design proposed in~\cite{zhu2010attacks} requires more than two minutes to generate a single challenge).

\paragraph{Attacks and countermeasures}
The attacks designed to automatically solve image-based CAPTCHAs are usually very specific, i.e., the attacker has to exploit weak points of each specific CAPTCHA design. 
The main attack strategies used against image-based CAPTCHAs are the following (to avoid confusion and have a unique numbering for attack strategies---also considering the ones for text-based CAPTCHAs---we continue from A08):
\begin{enumerate}[leftmargin=1.cm]
  \item[A08)]  Some CAPTCHAs (especially the ones based on games) hide the solution on client-side.
   Henceforth, an attacker might run what we call \textit{indirect attack}:  get the solution from the client-side (e.g., via reverse engineering of the client application).
  \item[A09)] Some CAPTCHAs rely on a pool of pre-computed challenges, stored in a database.
  A malicious attacker can perform the \textit{exhaustion of the database} using real humans (e.g., via Amazon Mechanical Turk\footnote{\url{https://www.mturk.com/}.}).
  \item[A10)] Similarly, an attacker can make queries to a \textit{leaked database} to identify the solution of a challenge.
  \item[A11)] An attacker can use \textit{machine learning} techniques (e.g., Support Vector Machine) to recognize the objects that compose a challenge and solve it. 
  \item[A12)] In case of a limited number of possible answers, an attacker could simply rely on a \textit{random chance} obtaining a decent success rate.
  \item[A13)] CAPTCHAs solvable with a single interaction are prone to \textit{pure relay attacks}. Indeed, attackers can simply send a screenshot of the challenge to an external paid human. 
  \item[A14)] Given a heavily interactive CAPTCHA, a bot can synchronously relay the data stream from the server over to a human solver, and then relay back the input of the user to the server.
  This strategy is defined as \textit{stream relay attack}~\cite{mohamed2014three}.
  
\end{enumerate}
Several improvements are possible to mitigate the previous weaknesses. Some examples follow: 
\begin{itemize} %
  \item Use code obfuscation or encryption (against A08). 
  \item Use Web crawlers to have a self-growing database (against A09). 
  \item Process objects stored in the database before presenting them in the challenge. 
  This makes it unfeasible 
  to match the original object with the one presented in the challenge (against A09 and A10). 
  \item Enlarge the search space in order to increase 
  the computational cost to find a solution (against A11). 
  \item Increase the number of possible answers (against A12). 
  \item Analyze the behavioral features, identifying suspicious pattern of movement~\cite{mohamed2014dynamic} (against A13 and A14). 
\end{itemize}
\section{Our proposal: CAPTCHaStar}
\label{ourproposed}
In this section, we present CAPTCHaStar, a novel image-based CAPTCHA.
The aim of our proposal is to provide a high level of usability, while improving security.
In the following, we first provide a high level overview of the system (Section~\ref{captchastaroverview}), then we discuss the actual implementation of the prototype (Section~\ref{prototypeimplementation}).

\subsection{CAPTCHaStar overview}
\label{captchastaroverview}
\begin{figure*}[ht!]
\centering
\begin{subfigure}{.24\textwidth}
 \centering
\includegraphics[width=0.95\textwidth]{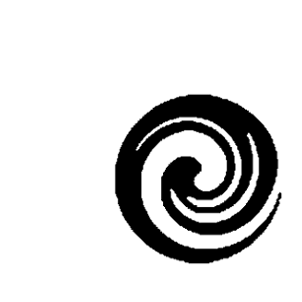}
\caption{A random starting picture.}
\label{fig:originSpritz}
\end{subfigure}
\begin{subfigure}{.24\textwidth}
  \centering
  \includegraphics[width=0.95\textwidth]{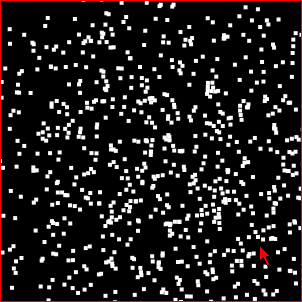}
  \caption{A sample unsolved challenge.}
\label{fig:unsolvedSpritz}
\end{subfigure}
\begin{subfigure}{.24\textwidth}
  \centering
  \includegraphics[width=0.95\textwidth]{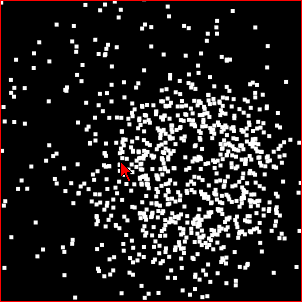}
  \caption{An almost solved challenge.}
\label{fig:almostSpritz}
\end{subfigure}
\begin{subfigure}{.24\textwidth}
  \centering
  \includegraphics[width=0.95\textwidth]{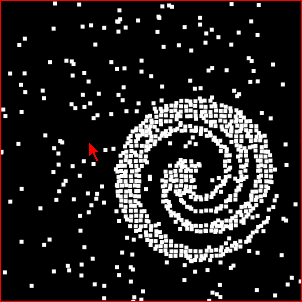}
 \caption{A correctly solved challenge.}
\label{fig:solvedSpritz}
\end{subfigure}
\vspace{-0.0cm}
\caption{The process of solving a CAPTCHaStar challenge.}
\label{fig:test}
\end{figure*}

Our CAPTCHA prompts the user with several small white squares, randomly placed inside a squared black space.
From now on, we refer to a single white square as a \textit{star}, and to the squared black space as the \textit{drawable space}.
The position of each star
changes according to
the current coordinates of the cursor, inside the drawable space.
Given a challenge, we define as \textit{state} a snapshot of the stars location on the drawable space, relative to a specific cursor position.
The challenge asks the user
(who wants to be recognized as a human)
to change the position of the stars,
by moving her cursor, until she is able to recognize a shape (which is not predictable).
In particular, CAPTCHaStar creates such a shape starting from a picture randomly chosen among a huge set of pictures.
Figure~\ref{fig:originSpritz} illustrates an example of a picture with ideal features: two colors and a limited number of small details.


Our system decomposes the selected picture in several stars using a sampling algorithm (described later in Section~\ref{prototypeimplementation}).
%
For each star, the system sets its movement pattern, in a way such that the stars can 
aggregate together, forming the shape of the sampled picture. This happens only when the cursor is on a secret position.
We refer to that position as the \textit{solution} of the challenge.
In general, a single CAPTCHaStar challenge can include more than one shape, each of them having its own solution (i.e., secret position of the cursor), at which becomes visible. 

%

When the position of the cursor is far from the solution, the stars appear randomly scattered on the black space.
Figure~\ref{fig:unsolvedSpritz} shows an example, obtained from the stars that compose the picture in Figure~\ref{fig:originSpritz}.
The user has to move the cursor inside the drawable space until she recognizes a meaningful shape.
As the distance between the cursor and the solution decreases significantly, the stars aggregate together
in a more and more detailed shape
(see Figure~\ref{fig:almostSpritz}).
The user needs to adjust the position of the cursor, until she is confident that the resulting shape is detailed enough (see Figure~\ref{fig:solvedSpritz}).
Finally, the user confirms the current cursor position as her final answer.
The system compares the solution with the final answer (allowing a small margin of error), eventually assessing whether the interaction was made by human.


%

To make the solution of the CAPTCHA more difficult for a bot, in addition to the stars forming the original shape (\textit{original stars}), we add also \textit{noisy stars}: i.e., stars that will be in random position when the shape is complete.
The number of the noisy stars can be tuned according to a specific parameter. 

The system stores on server-side the solution of the challenge, 
and performs the  
check only when the user confirms her answer, that is considered as final and irrevocable.
For the sake of usability,
CAPTCHaStar considers as a valid answer also a pair of coordinates that is close enough to the actual solution (more details in Section~\ref{prototypeimplementation}).

The generation phase of a challenge involves some parameters to tune usability and security: 
\label{generationparameters}
\begin{itemize} 
  \item \textit{Noise ($\psi$)}: the percentage of noisy stars added to the scheme, with respect to the number of original stars.
  \item \textit{Sensitivity ($\delta$)}:
  the relationship between the amount of displacement of the cursor (in pixel) and the movement of each star  
  (more details in Section~\ref{prototypeimplementation}). 
  \item \textit{NSol}: 
  the number of possible solutions (i.e., secret positions) of the challenge.
  Each solution corresponds to a different shape.
  \item \textit{PicSize}: the maximum value between width and height on the sampled picture, expressed in number of pixels. 
  \item \textit{Rotation}: Boolean value that indicates whether the picture is rotated by a random degree. 
\end{itemize}

\subsection{Prototype implementation}
\label{prototypeimplementation}
To assess the feasibility and effectiveness of our solution, we did a complete implementation.
In particular, we aimed at providing an implementation that could be widely deployed.
Since \textit{PHP} is the most supported programming language by web servers~\cite{commonserversidelanguage},
we implemented the server-side part of our design using this language.
We implemented the client-side part using \textit{HTML5 Canvas}, because it has the support for majority of commercial browsers~\cite{canvassupport}. 
We manually retrieved more than $5000$ pictures with two colors, by searching free-to-use collections\footnote{Free vector icons, \url{http://www.flaticon.com/}.} of vector icons in \texttt{.png} format (this step could be automated, e.g., with web crawlers).
We collected all these pictures in a pool. For a real life deployment of that system, we recommend using a pool as large as possible.
In the following, we first describe how a challenge is generated on the server-side, then we describe how it is presented to the user on the client-side.

\subsubsection{Generation of a challenge}
Each challenge is composed by original stars (generated from the base shape) and noisy stars (generated randomly).
The steps to generate a challenge are as follows:
i) Picture selection and pre-process; 
ii) Picture decomposition; 
	iii) Trajectory computation. 
Our system repeats these steps for a number of times equal to the value of the parameter \textit{NSol}.

\noindent\textit{Picture selection and pre-process.}
Our system randomly chooses one of the pictures from the pool, and resizes it according to the value of the parameter \textit{PicSize}.
If the \textit{Rotation} parameter is enabled, CAPTCHaStar rotates the picture by a random degree.
At this point, our system converts the picture in black and white (i.e., binarization). 

\noindent\textit{Picture decomposition.}
The sampling algorithm first divides the picture in $5$x$5$ pixel tiles, then it counts the number of black pixels inside each tile. 
A tile will result in an original star when 
it matches one of the following conditions:
(i) if the tile is filled with black pixels (i.e., having $5$x$5=25$ black pixels), our system generates an original star and places it at the center of the tile; 
(ii) if the tile has a number of black pixels between 9 and 24, our system generates an original star and places it in a position that is shifted from the center of the tile, toward the position where there are the majority of black pixels.
Our system places the final shape composed by stars inside the \textit{drawable space}, in a random position (such that all the original stars lie inside).

\noindent\textit{Trajectory computation.}
We define the solution $sol$ of the challenge as the pair of coordinates $({sol}_x,{sol}_y)$.
Our system generates ${sol}_x$ and ${sol}_y$ at random, within the range of $[5,295]$.
We adopted such range for the sake of usability.
In particular, this guarantees that the solution will not appear on the edges of the drawable area (which is $300$x$300$ pixel).
For each original star $i$, our system also defines $(P^i_x,P^i_y)$ as the coordinates of the position that the star $i$ takes when the cursor is in coordinates $({sol}_x,{sol}_y)$.  
For each star $i$, our system randomly generates four coefficients ($m^i_{x,x}$, $m^i_{x,y}$, $m^i_{y,x}$, $m^i_{y,y}$),
that relates the coordinates of the star with the coordinates of the cursor: 
$m^i_{ab}$ associates the coordinate of the star $i$ in axis $a$, with the coordinate of the cursor in axis $b$.
The values of these coefficients are picked in the range $[-\frac{\delta}{10}$, $\frac{\delta}{10}]$ (we remind that $\delta$ is the \textit{sensitivity} value).
Our system computes a pair of constants, ($C^i_x$, $C^i_y$), for each original star $i$ as follows:
\begin{gather*}
C^i_x = P^i_x-{sol}_y\cdot{m^i_{x,y}}-{sol}_x\cdot{m^i_{x,x}},\\
C^i_y = P^i_y-{sol}_y\cdot{m^i_{y,y}}-{sol}_x\cdot{m^i_{y,x}}.
\end{gather*}
CAPTCHaStar generates the noisy stars in a similar way, but their coordinates $(P_x,P_y)$ having random values. 
The number of noisy stars is equal to the percentage $\psi$ of the number of original stars.
Henceforth, we define as \textit{trajectories parameters} of star $i$, the following set of parameters: $m^i_{x,x}$, $m^i_{x,y}$, $C^i_x$, $m^i_{y,x}$, $m^i_{y,y}$, $C^i_y$.
The only information that the client needs from the server in order to calculate the position of the stars, whenever the user moves her cursor, is the trajectories parameters.
We underline that noisy and original stars are mixed together, i.e., they are indistinguishable from client side. 
\subsubsection{Presentation of a challenge}
Whenever the user moves the cursor, our system uses the cursor coordinates $cur=(cur_x, cur_y)$ to compute the new coordinates of each star $i$, as follows:
\begin{gather*}
x^i=m^i_{x,y}\cdot {cur_y}+m^i_{x,x}\cdot {cur_x}+C^i_x,\\
y^i=m^i_{y,x}\cdot {cur_x}+m^i_{y,y}\cdot {cur_y}+C^i_y.
\end{gather*}
When the user 
confirms her answer (e.g., with a mouse click), the client passes $cur$ to a simple server-side script, via \texttt{HTTP GET} parameter.
For the sake of usability, on mobile devices the submission of the answer is performed by tapping on a button, which is external to the drawable space.

Our server-side script calculates $\Delta$ as the euclidean distance between $sol$ and $cur$.
We define \textit{usability tolerance} as a threshold, in terms of euclidean distance from $sol$. 
When the value $\Delta$ is below the \textit{usability tolerance}, the system considers the test as passed (failed otherwise).  
From our experiments, we found that a reasonable value for \textit{usability tolerance} is close to five \hilight{(more details in Section}~\ref{parametersselection}). 
We highlight that the position of each star varies linearly with the movement of the cursor.
For this reason, humans can easily build a mental map~\cite{norman2013design} 
of the stars' behavior, hence moving the cursor toward the position that is closer to a real shape.

\subsection{System performance}
\begin{figure}[ht]
\centering
\includegraphics[width=0.48\textwidth]{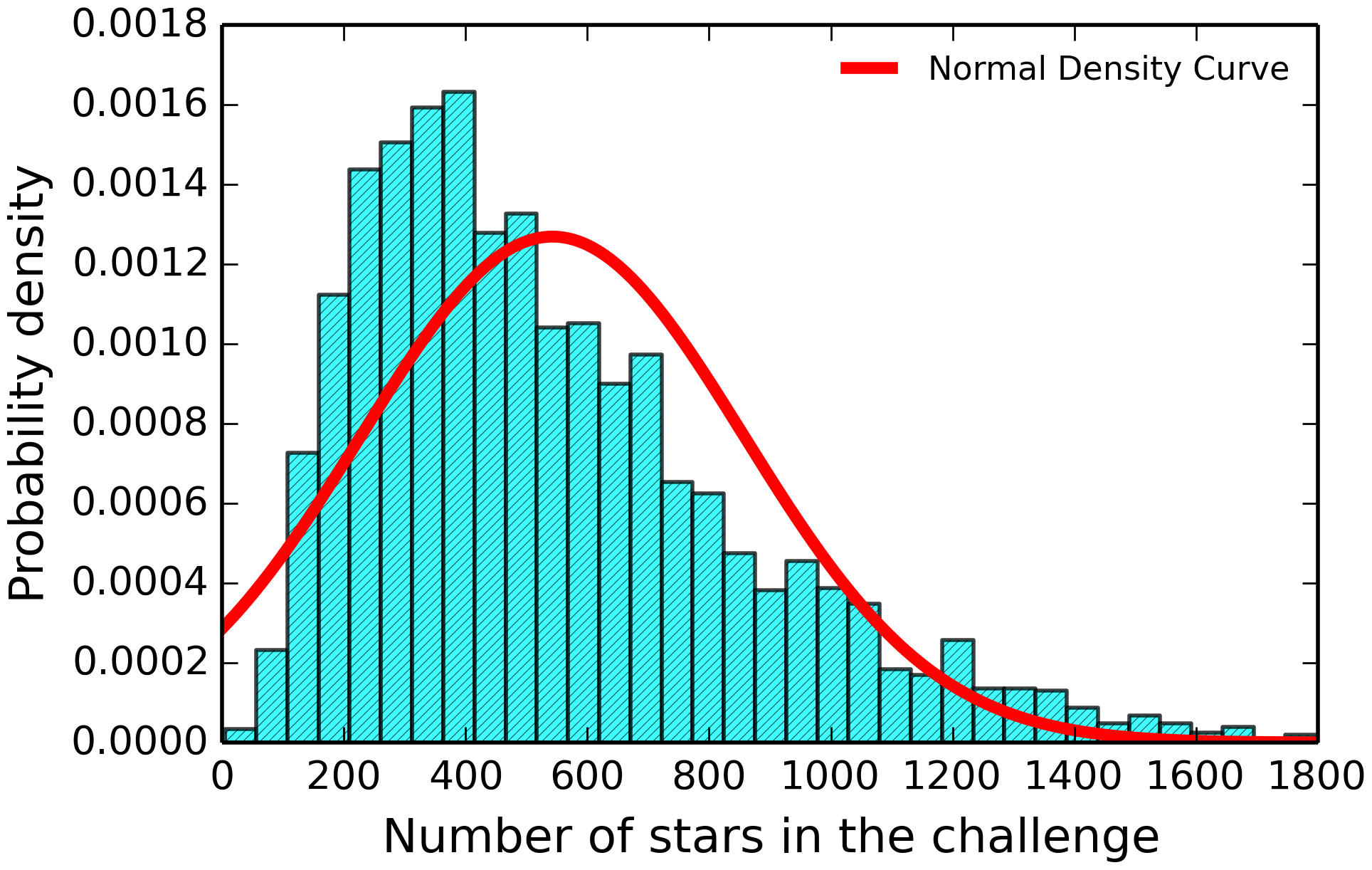}
\caption{Probability distribution of the number of stars.}
\label{fig:histGeneration}
\end{figure}

\begin{figure}[ht]
\centering
\includegraphics[width=0.48\textwidth]{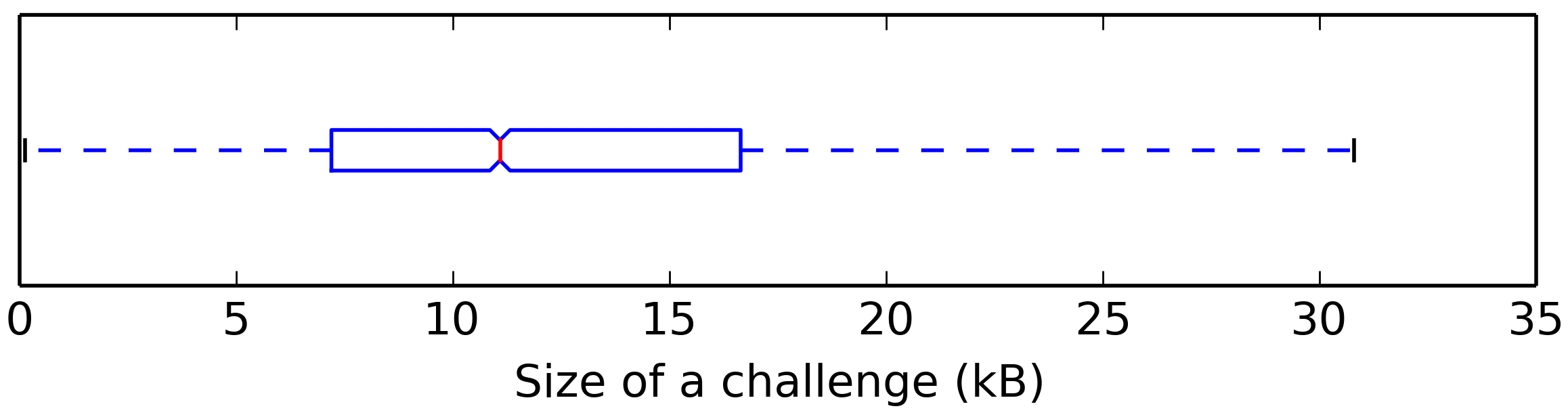}
\caption{Statistical distribution of the sizes of the challenges in kilobytes. The
notch of the box represents the median value. First and third quartile
are represented as the left and right side of the notched box.  
Lines that extend horizontally from the boxes indicate the $2^{nd}$ percentile (left) and
the $98^{th}$ percentile (right).}
\label{fig:sizeChallenge}
\end{figure}

\hilight{In this section, we discuss the performance of our proposal in terms of computational time and bandwidth.
The performance is directly related to the number of stars in a challenge.}
In Figure~\ref{fig:histGeneration}, \hilight{we report the probability density of the number of stars in a challenge, computed from our database with $4033$ challenges.
This probability density results in a Normal Density Curve with mean $\mu = 542.7$ and standard deviation $\sigma = 314.4$.}
According with this finding, for a CAPTCHaStar challenge with an average number of stars (i.e., 543~stars), the server sends to the client $12.7$~kB (i.e., 4~bytes for each one of the six parameters).
In Figure~\ref{fig:sizeChallenge}, we report the distribution of network overheads caused by the request of a CAPTCHaStar challenge. 
We can observe that more of the $75\%$ of the challenges are smaller than $17$~kB.
\hilight{We underline that in out current prototype implementation, the parameters are passed without any compression, while a compression could reduce the amount of Bytes transmitted.
As a comparison with the existing CAPTCHAs, we point out that a text-based reCaptcha challenge sends some $3$~kB, while an image-based reCaptcha challenge sends some $35$~kB. 
As an additional example, a single Asirra challenge needs more than $360$~kB.}

\begin{figure}[ht]
\centering
\includegraphics[width=0.48\textwidth]{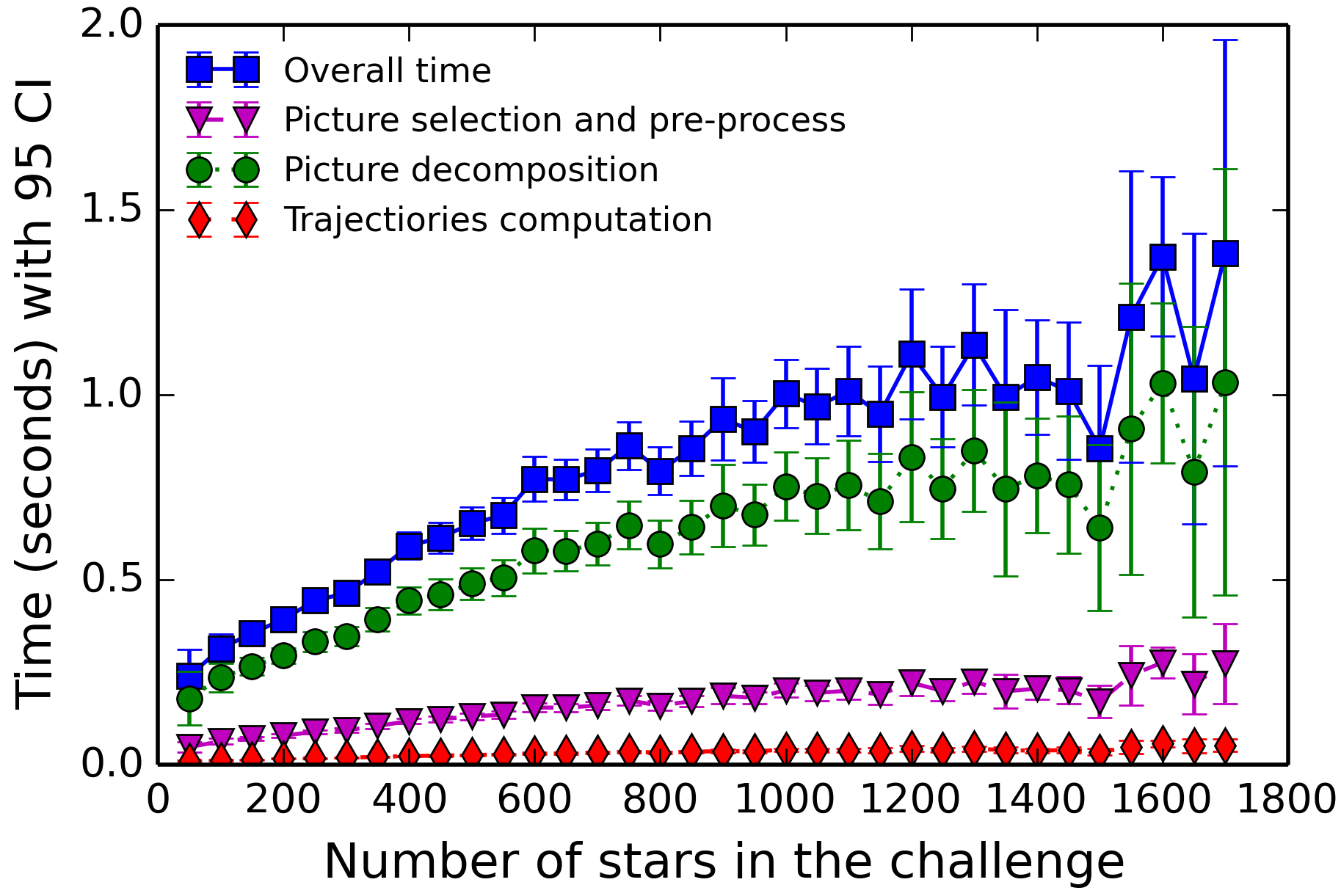}
\caption{Challenge generation time.}
\label{fig:timesGeneration}
\end{figure}

The server we use to generate the challenges is a PC with a $3.0$~GHz AMD Athlon Dual Core Processor and $1$~GB of RAM.
In Figure~\ref{fig:timesGeneration},
\hilight{we report the average time required by each phase of our solution to generate a challenge described above, depending on the number of stars.
As we can see from Figure~\ref{fig:timesGeneration}, the most costly phase of the process is the \textit{picture decomposition} phase, i.e., the generation of the original stars by sampling the starting image.
This phase takes some $76\%$ of the overall time to generate a challenge.
The \textit{picture selection and pre-process} phase (i.e., loading, rotating and resizing the image) requires around $21\%$ of the overall time.
Only some $3\%$ of the overall challenge generation time is due to the \textit{trajectory computation} phase.
Considering the average challenge (i.e., 542~stars), the overall time required to generate such challenge is about $0.75$~seconds.
We underline that even in the worst case scenario, the overall time is always lower than two seconds.}
This suggests that our prototype can handle efficiently a wide number of requests, even with low cost hardware resources.
\hilight{Moreover, we believe the generation time would in no way be a showstopper for CAPTCHAs (and for our proposal in particular), since a possible solution (indeed applicable to several CAPTCHAs) would be either to generate the challenge while the user is doing other operations, or to maintain a pool of pre-generated challenges, and randomly pick one when needed.}

%

\subsection{User interaction}
\label{user_interaction}
\begin{figure}[h!]
\centering
\begin{subfigure}{.22\textwidth}
\centering
\includegraphics[width=0.98\textwidth]{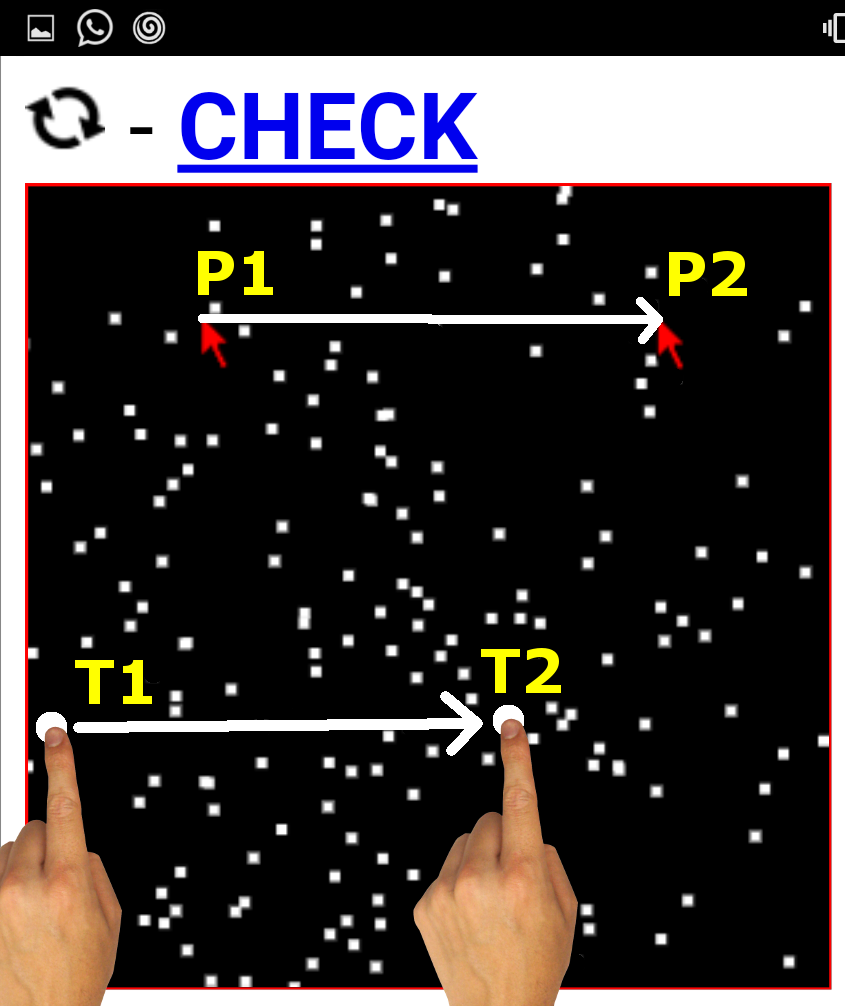}
\caption{First interaction.}
\label{fig:mobileInterface1}
\end{subfigure}
\begin{subfigure}{.22\textwidth}
\centering
\includegraphics[width=0.98\textwidth]{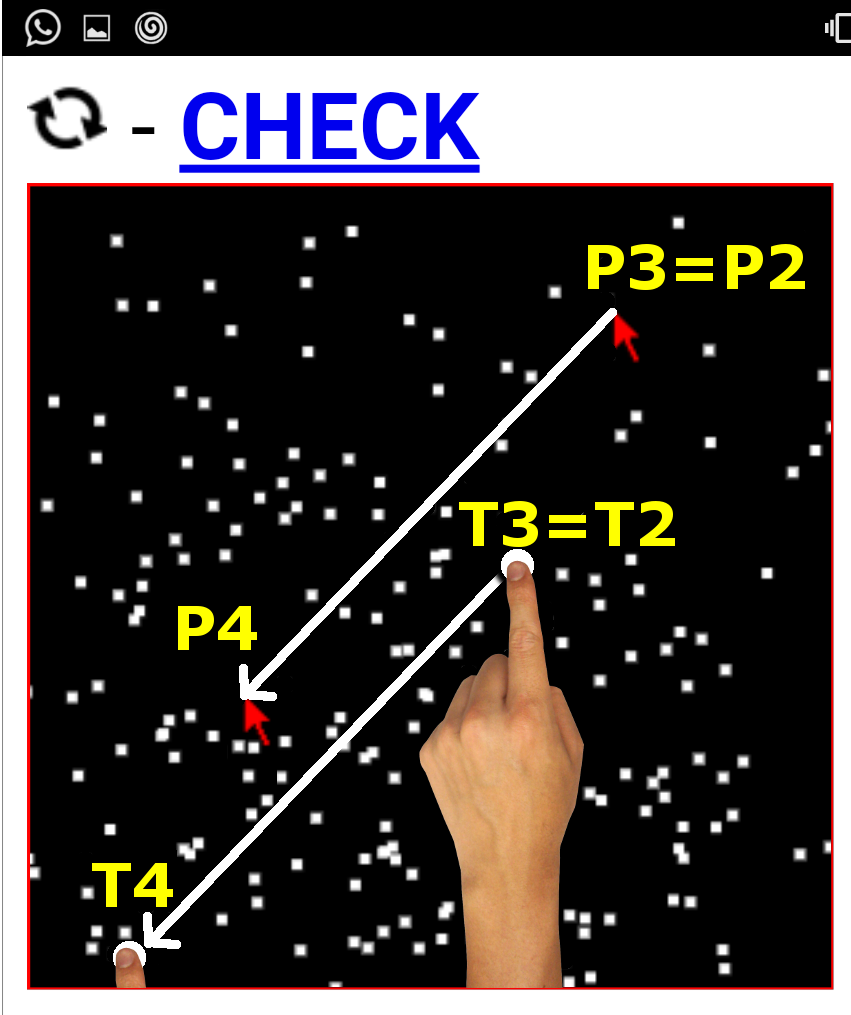}
\caption{Second interaction.}
\label{fig:mobileInterface2}
\end{subfigure}
\vspace{-0.25cm}
\caption{User interaction examples on touchscreen devices.}
\label{fig:mobileInterface}
\end{figure}

For the sake of usability of CAPTCHaStar, we consider different cursors according to the device it runs on. 
In particular, considering a browser on a personal computer, the cursor coincides with the default mouse pointer, and the user can submit her final answer by clicking the left mouse button. 
On the other hand, on smartphones or other touch-enabled devices, the cursor position is usually not represented with a graphical object, such as the mouse pointer on personal computers.
For this reason, we represent the cursor position as a red arrow inside the drawable area (as shown in Figure~\ref{fig:mobileInterface}).
The user can move the cursor by swiping her finger on the drawable area (note that only the starting point of the swipe must be inside the drawable area). 
In Figure~\ref{fig:mobileInterface1}, we report a first example of a user moving the cursor from position P1 to P2, by swiping the finger from T1 to T2 on the touchscreen. 
In Figure~\ref{fig:mobileInterface2}, we show a second example, where the starting position P3 of the cursor corresponds to the ending position P2 in Figure~\ref{fig:mobileInterface1}. In the second example, a user moves the cursor from position P3 to P4, with a swipe from T3 to T4. 
We highlight that the path of the cursor is mapped directly to the path of the swipe, regardless of the starting point of that swipe.
Moreover, the position of the cursor at the end of the swipe in Figure~\ref{fig:mobileInterface1} (i.e., T2) remains in position P2. 

In order to submit her final answer on a mobile browser, the user has to tap on a ``CHECK'' link placed outside of the drawable area, as shown at the top of figures~\ref{fig:mobileInterface1} and~\ref{fig:mobileInterface2}.

\section{User study} 
\label{evaluateusability}
In order to evaluate our proposal, we ran a user study according to the usability metrics proposed in~\cite{yan2008usability}, and an exhaustive set of parameter combinations.
In particular, we compare our solution with text-based CAPTCHAs taken from reCaptcha~\cite{recaptchaplugin}. 
In the following, we describe in detail how we ran the user study and discuss the obtained results.  
%
\subsection{Parameters selection}
\label{parametersselection}
In order to notify the user whether she passes or fails a challenge, we need to set the value of the \textit{usability tolerance} parameter (introduced in Section~\ref{prototypeimplementation}).
On one hand, increasing the value of this parameter makes CAPTCHaStar more permissive.
On the other hand, it also increases the success rate of random guess attacks with a quadratic growth.
For this reason, it was crucial to identify an optimal value for the \textit{usability tolerance} parameter, in order to obtain a good trade-off between usability and security.

We ran this preliminary study on a set of $35$ participants (volunteers and without any reward) under our supervision. 
A session of this study consisted in solving three CAPTCHaStar tests (named P1, P2 and P3) for at least two times (i.e., six challenges in total).
\hilight{CAPTCHaStar tests are randomly generated using the parameter values reported in Table}~\ref{tab:testpreliminar}. 

\begin{table}[ht]
\centering
\scalebox{1.}{
\begin{tabular}{|l|c|c|c|c|c|c|} \hline
\textbf{Test} & \textbf{P1} & \textbf{P2} & \textbf{P3} \\ \hline
\textbf{$\boldsymbol\psi$} & $0\%$ & $70\%$ & $70\%$ \\ \hline
\textbf{$\boldsymbol\delta$} & $5$ & $7$ & $7$ \\ \hline
\textbf{NSol} & $1$ & $1$ & $1$  \\ \hline
\textbf{Rotation} & Off & Off & On \\ \hline
\end{tabular}
}
\caption{Values of parameters $\psi$, $\delta$, \textit{NSol} and \textit{Rotation} for the preliminary study.}
\label{tab:testpreliminar}
\end{table}

\begin{figure}[t]
\centering
\includegraphics[width=0.48\textwidth]{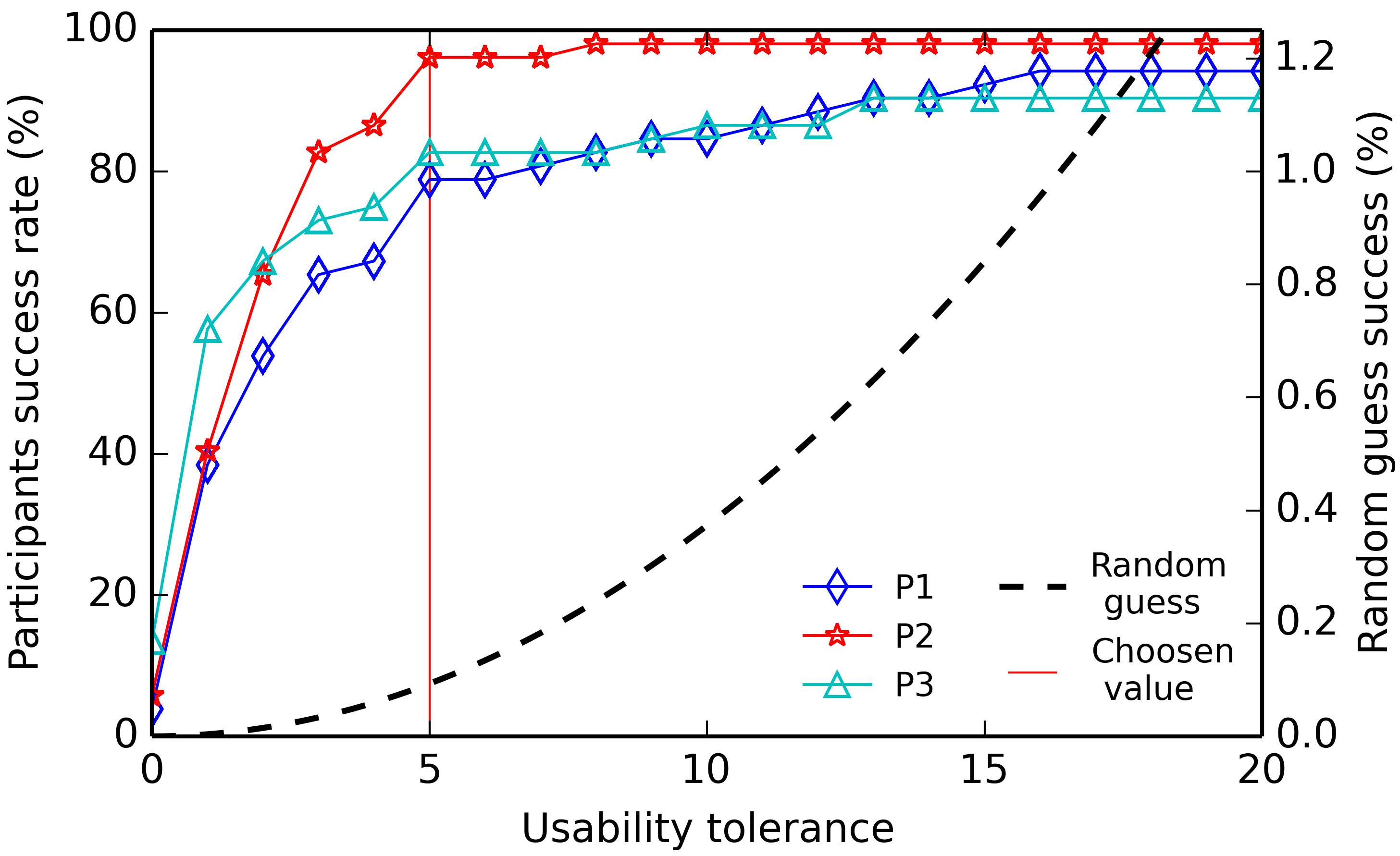}
\caption{Success rate of the participants of the preliminary study (scale on the left-hand side of the graph) and success rate of random guess attack (scale on the right-hand side of the graph), varying the value of \textit{usability tolerance}.}
\label{fig:usability_tollerance_first}
\end{figure}

Figure~\ref{fig:usability_tollerance_first} reports the success rate of the participants and success rate of random guess attacks, as the \textit{usability tolerance} varies.
In particular, 
we noticed that the participants success rate grows rapidly until \textit{usability tolerance} is equal to five, then it plateaus.
Since the random guess attack success rate grows very fast (i.e., has a quadratic growth), and the user success rate does not improve significantly with a \textit{usability tolerance} greater than five, for our user study (described in the following) we set the \textit{usability tolerance} equal to five.

\subsection{Survey design and implementation}
\begin{figure}[h]
\centering
\begin{subfigure}{.22\textwidth}
\centering
\includegraphics[width=0.99\textwidth]{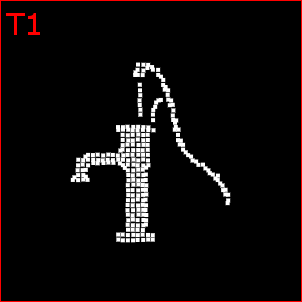}
\caption{T1}
\label{fig:T1}
\end{subfigure}
\begin{subfigure}{.22\textwidth}
\centering
\includegraphics[width=0.99\textwidth]{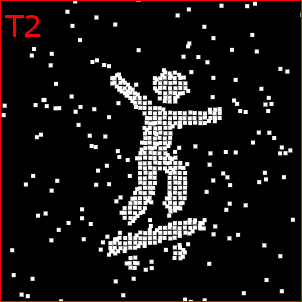}
\caption{T2}
\label{fig:T2}
\end{subfigure}

\begin{subfigure}{.22\textwidth}
\centering
\includegraphics[width=0.99\textwidth]{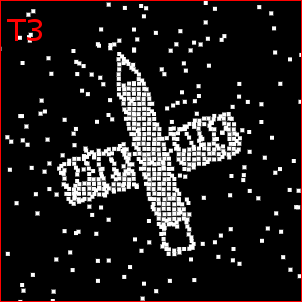}
\caption{T3}
\label{fig:T3}
\end{subfigure}
\begin{subfigure}{.22\textwidth}
\centering
\includegraphics[width=0.99\textwidth]{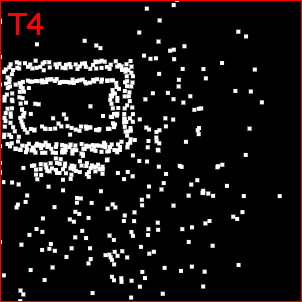}
\caption{T4}
\label{fig:T4}
\end{subfigure}

\begin{subfigure}{.22\textwidth}
\centering
\includegraphics[width=0.99\textwidth]{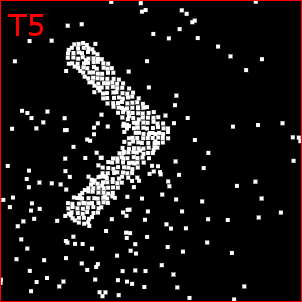}
\caption{T5}
\label{fig:T5}
\end{subfigure}
\begin{subfigure}{.22\textwidth}
\centering
\includegraphics[width=0.99\textwidth]{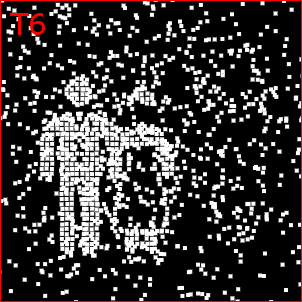}
\caption{T6}
\label{fig:T6}
\end{subfigure}
\caption{Screenshots of examples of solved challenges, generated randomly using the settings from T1 to T6.}
\label{fig:examples_captchas}
\end{figure}

We designed a web-based survey page, in order to collect data from a large number of participants.
We built a survey composed of eight different tests: six CAPTCHaStar challenges (named from T1 to T6) and two text-based ones (T7 and T8).
Tests from T1 to T6 are randomly generated (i.e., starting from a random image) using the value of parameters reported in Table~\ref{tab:testfeatures}. 
Tests T4 and T5 have more than one solutions, i.e., two and three, respectively. Test T4 requires the user to find both of its solutions, while for T5, it is enough to find only one of the three existing solutions.
In Figure~\ref{fig:examples_captchas}, we present some examples of solved challenges for the settings from T1 to T6. 
%
%
\begin{table}[t!]
\centering
\scalebox{1.}{
\begin{tabular}{|>{\hspace{\tabred}}l<{\hspace{\tabred}}|
		>{\hspace{\tabred}}c<{\hspace{\tabred}}|
		>{\hspace{\tabred}}c<{\hspace{\tabred}}|
		>{\hspace{\tabred}}c<{\hspace{\tabred}}|>{\hspace{\tabred}}c<{\hspace{\tabred}}|
		>{\hspace{\tabred}}c<{\hspace{\tabred}}|
		>{\hspace{\tabred}}c<{\hspace{\tabred}}|}
\hline
\textbf{Test} & \textbf{T1} & \textbf{T2} & \textbf{T3} & \textbf{T4} & \textbf{T5} & \textbf{T6} \\ \hline
\textbf{$\boldsymbol\psi$} & $0\%$ & $70\%$ & $70\%$ & $10\%$ & $0\%$ & $250\%$ \\ \hline
\textbf{$\boldsymbol\delta$} & $5$ & $7$ & $7$ & $7$ & $10$ & $5$ \\ \hline
\textbf{NSol} & $1$ & $1$ & $1$ & $2$ & $3$ & $1$ \\ \hline
\textbf{Rotation} & Off & Off & On & Off & Off & Off \\ \hline
\textbf{Usability tolerance} & 5 & 5 & 5 & 5 & 5 & 5 \\ \hline
\end{tabular}
}
\caption{Values of parameters $\psi$, $\delta$, \textit{NSol}, \textit{Rotation} and  \textit{Usability tolerance} for the survey.}
\label{tab:testfeatures}
\end{table}

%
The last two tests are \hilight{random} text-based CAPTCHAs from reCaptcha, with one and two words (i.e., T7 and T8, respectively). 
In order to minimize the learning effect~\cite{kosara2003user}, 
we prompt the user with the eight tests selected in a random order.
At the beginning of the survey, we prompt users with a description of our proposal and a simple demo.
Then, we ask the participants to fill out a form with their demographic information: age, gender, nationality, level of education, years passed using Internet, and frequency of Internet use. 
We gather this data in order to understand whether factors like the experience of the user affects the performances in solving CAPTCHaStar challenges.
In the same page, we also ask the participants to read and accept an informed consent statement, where we declare how we intend to use the collected data and that we do not intend to disclose private information to third parties.
\hilight{For each test in the survey, we ask the user to rate the perceived difficulty of that test on a scale from 1 to 5.}
At the end of the eighth test, we asked the participants to: (i) rate the ease of understanding; (ii) indicate if they prefer our proposal or text-based CAPTCHAs; (iii) leave us any suggestion.
We design this survey in a way that each session should last less than 10 minutes. 
%
%
\subsection{Participants}
All the participants took the survey unsupervised 
using their own devices, in order to recreate the natural conditions of use of CAPTCHaStar.
We recruited the participants with an invitation (including a public link to the survey) that we broadcast on mailing lists and on social networks (i.e., Facebook, Google+, Twitter, and LinkedIn), in order to collect usage data for a large number of participants.
We did not give any reward for the participation.
More than $250$ users took part in our survey (\hilight{258 users, }$81\%$ male and $19\%$ female). 
We made sure that none of the $35$ participants of the preliminary study took part to this user study.
The average age of the participants was $25.5$. 
The education level 
was distributed as follows: $32\%$ high school diploma, $29\%$ bachelor degree, $26\%$ master degree, $9\%$ PhD, and $4\%$ none of the previous ones.
The totality of the participants used Internet daily,  $49\%$ from 5 to 10 years, $33\%$ for more than 10 to 15 years, $28\%$ for more than 15 years.
The majority (some $90\%$) of the participants were Italians. 
\hilight{However, we did not notice significant differences in the performance of users of different nationalities}. 
\hilight{Finally, only a few participants used mobile devices (16 users), with performances similar to desktop users.}

\subsection{Results and discussion}
Among all the participants, only $35\%$ of them preferred traditional CAPTCHAs rather than CAPTCHaStar.
Table~\ref{tab:surveyresult} reports the success rate and the average solving time for each of the eight challenges described above. 

\begin{figure}[t!]
\centering
\begin{subfigure}{.48\textwidth}
\includegraphics[width=.9\textwidth]{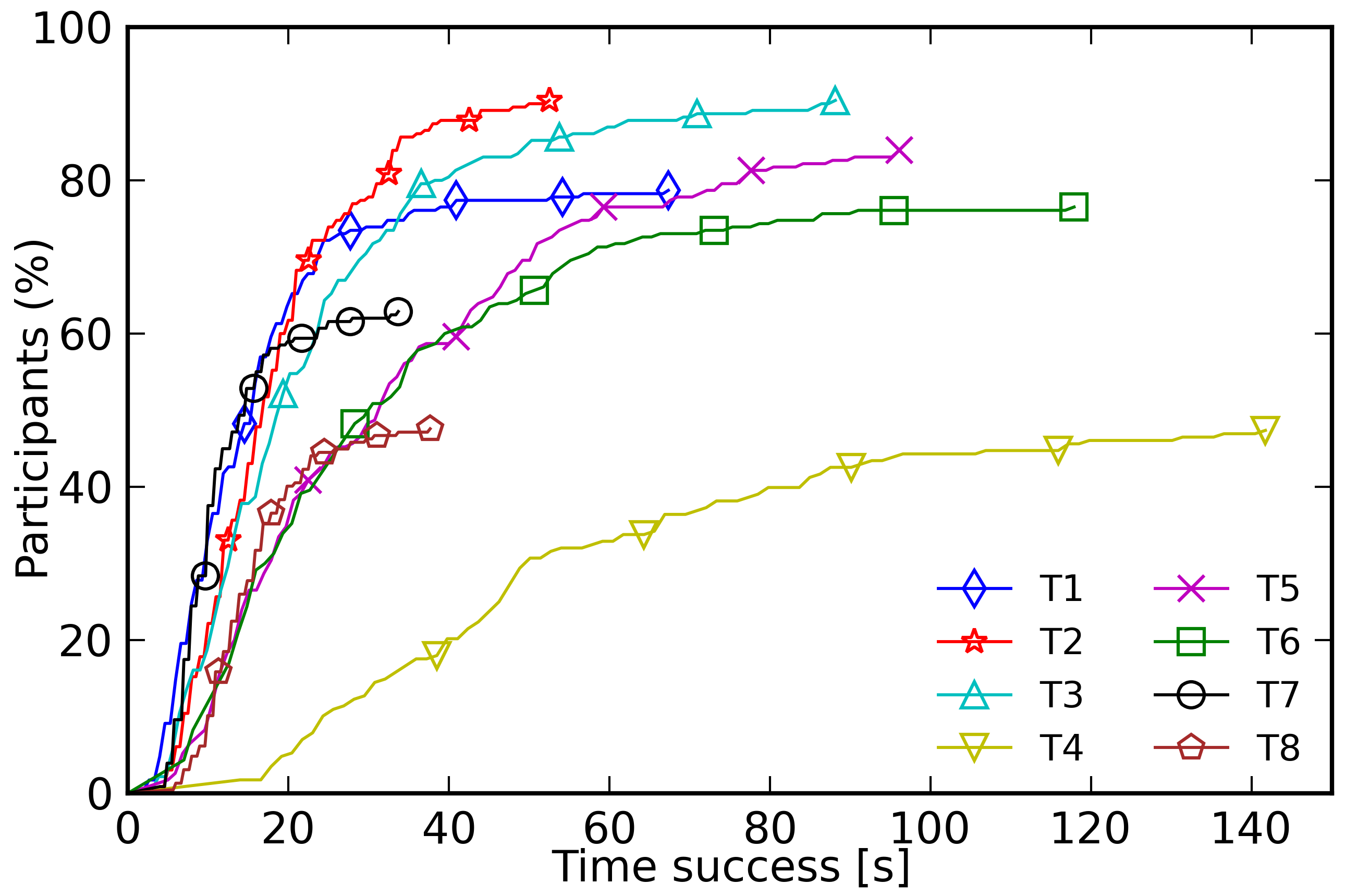}
\caption{Success rates over time.}
\label{fig:cdf_time_success}
\end{subfigure}
\\
\vspace{0.5cm}
\begin{subfigure}{.48\textwidth}
\includegraphics[width=.9\textwidth]{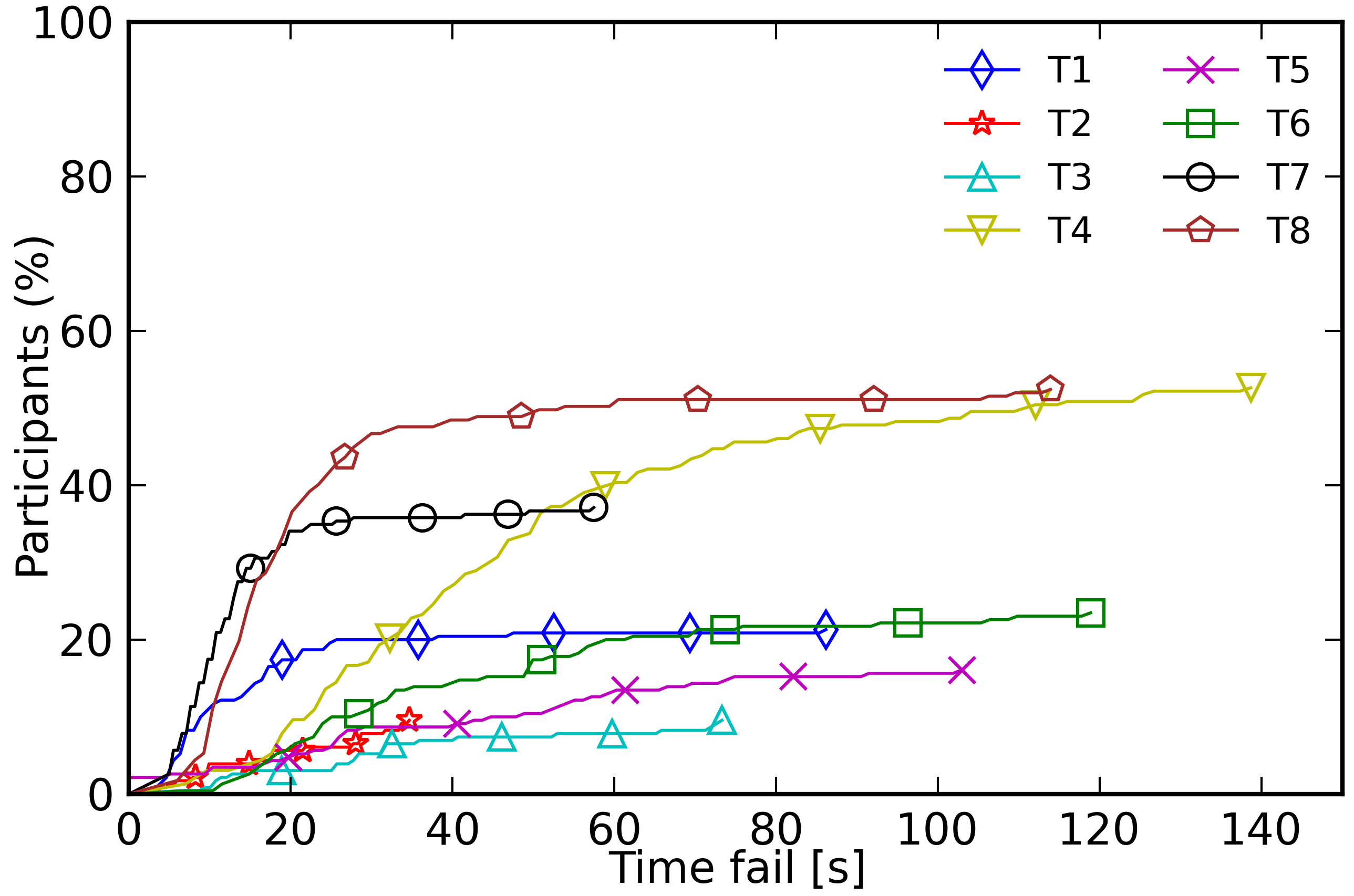}
\caption{Fail rates over time.}
\label{fig:cdf_time_fail}
\end{subfigure}
\caption{Success and fail rate over time. }
\label{fig:time_cdf}
\end{figure}

\begin{table}[h!]
\centering
\scalebox{1.}{
\begin{tabular}{|>{\hspace{\tabred}}c<{\hspace{\tabred}}|
		>{\hspace{\tabred}}c<{\hspace{\tabred}}|
		>{\hspace{\tabred}}c<{\hspace{\tabred}}|
		>{\hspace{\tabred}}c<{\hspace{\tabred}}|
		>{\hspace{\tabred}}c<{\hspace{\tabred}}|
		>{\hspace{\tabred}}c<{\hspace{\tabred}}|
		>{\hspace{\tabred}}c<{\hspace{\tabred}}|
		>{\hspace{\tabred}}c<{\hspace{\tabred}}|
		>{\hspace{\tabred}}c<{\hspace{\tabred}}|
		>{\hspace{\tabred}}c<{\hspace{\tabred}}|} \hline

\multicolumn{2}{|>{\hspace{\tabred}}c<{\hspace{\tabred}}|}{} & \multicolumn{6}{>{\hspace{\tabred}}c<{\hspace{\tabred}}}{\textbf{CAPTCHaStar}}&\multicolumn{2}{|>{\hspace{\tabred}}c<{\hspace{\tabred}}|}{\textbf{Text}}\\ \hline
\multicolumn{2}{|>{\hspace{\tabred}}c<{\hspace{\tabred}}|}{\textbf{Test}} & \textbf{T1} & \textbf{T2} & \textbf{T3} & \textbf{T4} & \textbf{T5} & \textbf{T6} & \textbf{T7} & \textbf{T8} \\ \hline
\multicolumn{2}{|>{\hspace{\tabred}}c<{\hspace{\tabred}}|}{\textbf{Succ. Rate (\%)}} & $78.7$ & $90.2$ & $90.6$ & $50.4$ & $85.1$ & $76.6$ & $62.7$ & $46.9$  \\ \hline
\multicolumn{2}{|>{\hspace{\tabred}}c<{\hspace{\tabred}}|}{\textbf{Difficulty}} & $1.9$ & $2.4$ & $2.6$ & $3.4$ & $2.9$ & $3.1$ & $2.4$ & $2.7$  \\ \hline \hline
\multirow{2}{*}{\textbf{Succ.}} & \textbf{Avg (s)} & $14.4$ & $17.5$ & $22.2$ & $54.1$ & $30.2$ & $28.5$ & $11.0$ & $14.9$  \\ \cline{2-10}
& \textbf{Std} & $9.8$ & $9.3$ & $15.8$ & $33.5$ & $20.2$ & $19.7$ & $5.4$ & $6.1$  \\ \hline
\multirow{2}{*}{\textbf{Fail}} & \textbf{Avg (s)} & $14.7$ & $18.2$ & $33.1$ & $49.0$ & $38.8$ & $40.0$ & $12.6$ & $21.2$  \\ \cline{2-10}
& \textbf{Std} & $13.5$ & $10.7$ & $21.2$ & $33.5$ & $26.6$ & $25.6$ & $8.8$ & $17.8$  \\ \hline
\end{tabular}
}
\caption{Survey results for CAPTCHaStar and text-based (Text) CAPTCHAs.}
\label{tab:surveyresult}
\end{table}
%
%
In most cases, when considering failed tests, the average completion time is higher than successfully passed ones.
%
In general, the standard deviation of these completion times is quite high (more than $25$ for most of the tests): a possible reason for this could be users having different abilities in solving CAPTCHaStar challenges. 
We highlight that all the CAPTCHaStar tests (i.e., T1 to T6) have a success rate higher than the one of T8 (i.e., text-based with two words), and only for T4 the success rate is lower than the one of T7 (i.e., text-based with one word).
We believe that users found T4 more difficult to be solved because it requires to discover two images (i.e., original stars for two images, plus the \textit{noisy stars}).
In particular, T2 shows a success rate that is some $90\%$, which is higher than the $84\%$ for text-based CAPTCHAs reported in~\cite{bursztein2010good}.
We underline that in our text-based CAPTCHAs T7 and T8 (where we used current reCaptcha used by Google), we observed a success rate of $62.7\%$ (for the simpler test with only one word).
In figures~\ref{fig:cdf_time_success} and~\ref{fig:cdf_time_fail}, 
we report in the domain of time the percentages of the participants that solved and failed a challenge, respectively. 

%
We highlight that text-based challenges (T7 and T8) rapidly approach 
to their maximum within some $20$ seconds, while CAPTCHaStar challenges reach a higher success rate in just a few more seconds. 
Indeed, the average time to solve T2 is some $17$ seconds, which is some $6$ seconds higher than the best time for text-based CAPTCHAs (i.e., $11$ seconds for T7).
We believe that this is an acceptable value.

\begin{figure}[ht]
\centering
\includegraphics[width=0.48\textwidth]{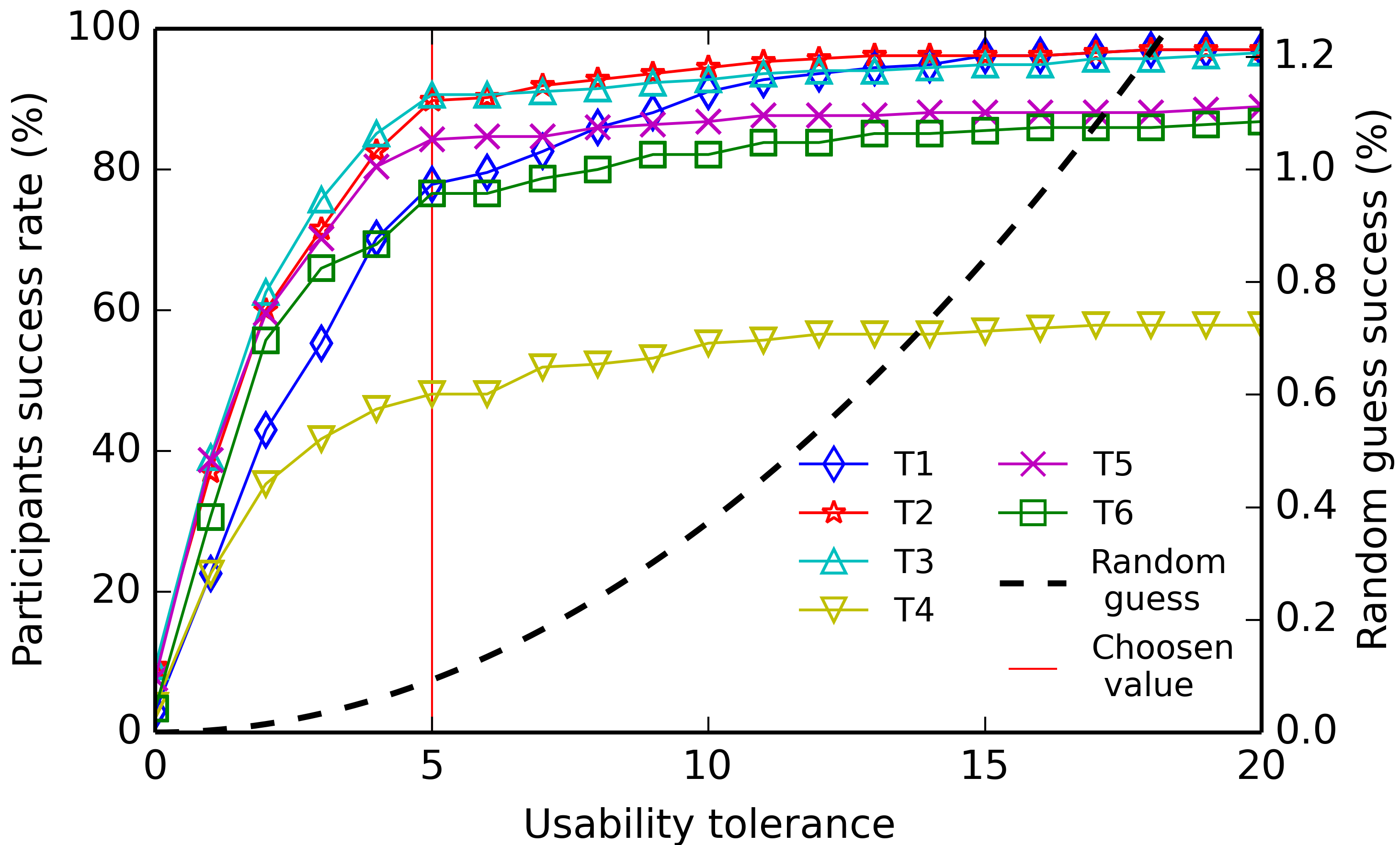}
\caption{Success rate of the participants of the main user study (scale on the left-hand side of the graph) and success rate of random guess attack (scale on the right-hand side of the graph), varying the value of \textit{usability tolerance}.}
\label{fig:usability_tollerance_second}
\end{figure}

In order to validate the choice of the \textit{usability tolerance} value, we report in Figure~\ref{fig:usability_tollerance_second} \hilight{a chart that now takes into account the data obtained from the main user study.}
This chart confirms our choice of five as being a reasonable value of \textit{usability tolerance}.
In fact, increasing the \textit{usability tolerance} (which also means giving more chances to adversaries) to a value above 5, brings only a limited benefit in the participants success rate -- see how the slope of most curves varies at \textit{usability tolerance} equal to 5.

Finally, we asked users to rate the ``ease of understanding'' \hilight{of the system} on a scale from 1 (very simple) to 10 (very difficult), and the results show an average value of $4.53$, with standard deviation of $2.53$.
In general, the comments received from the participants were positive. In particular, more than one gave comments similar to the following: ``I think CAPTCHaStar is better than text-based CAPTCHAs because it's a kind of game''. Other participants said something like ``text-based CAPTCHAs may require less time than yours, but I prefer CAPTCHaStar because I actually enjoyed it''.

\subsection{Learning effect evaluation}
\begin{figure}[t]
\centering
\begin{subfigure}{.48\textwidth}
\includegraphics[width=0.98\textwidth]{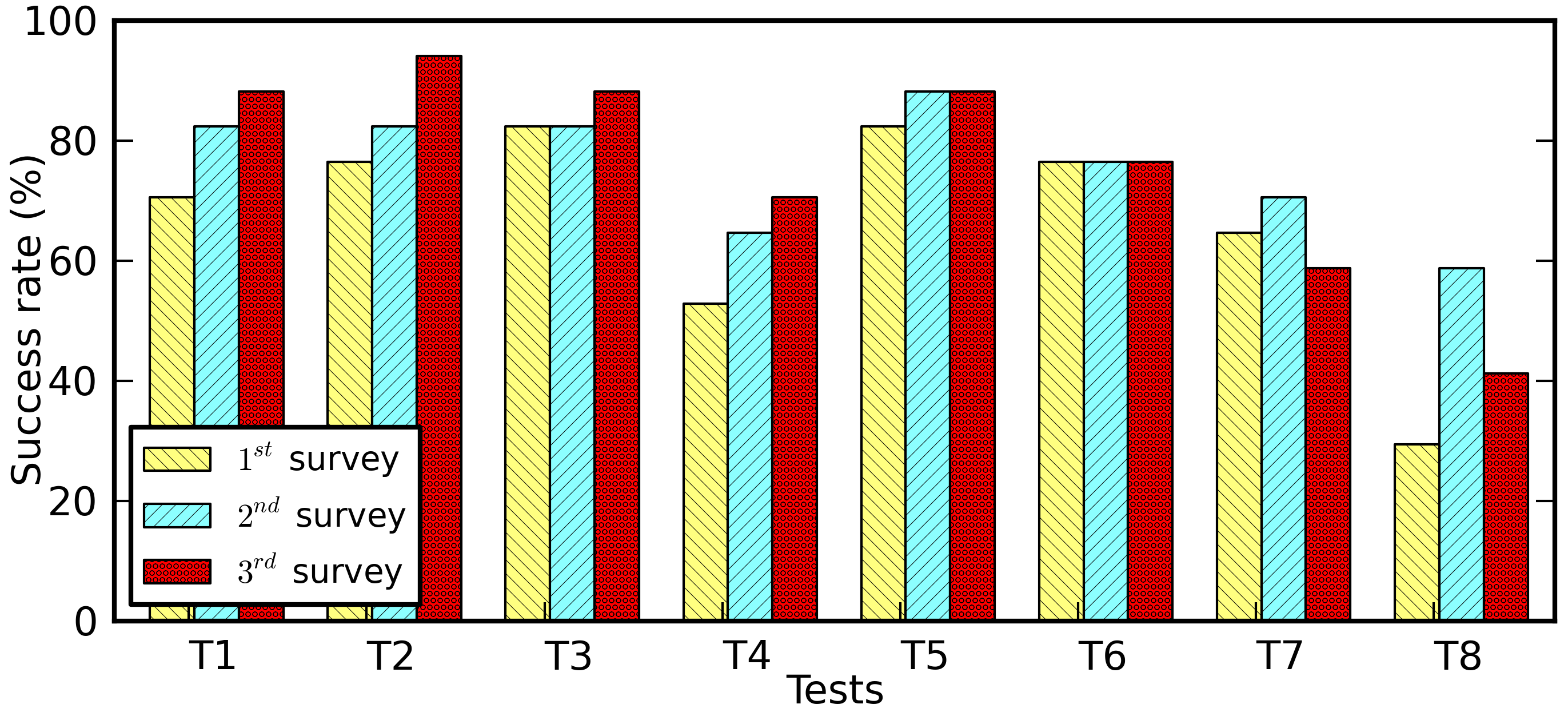}
\caption{Success rate.}
\label{fig:learning_effect_rate}
\end{subfigure}
\\
\vspace{0.5cm}
\begin{subfigure}{.48\textwidth}
\includegraphics[width=0.98\textwidth]{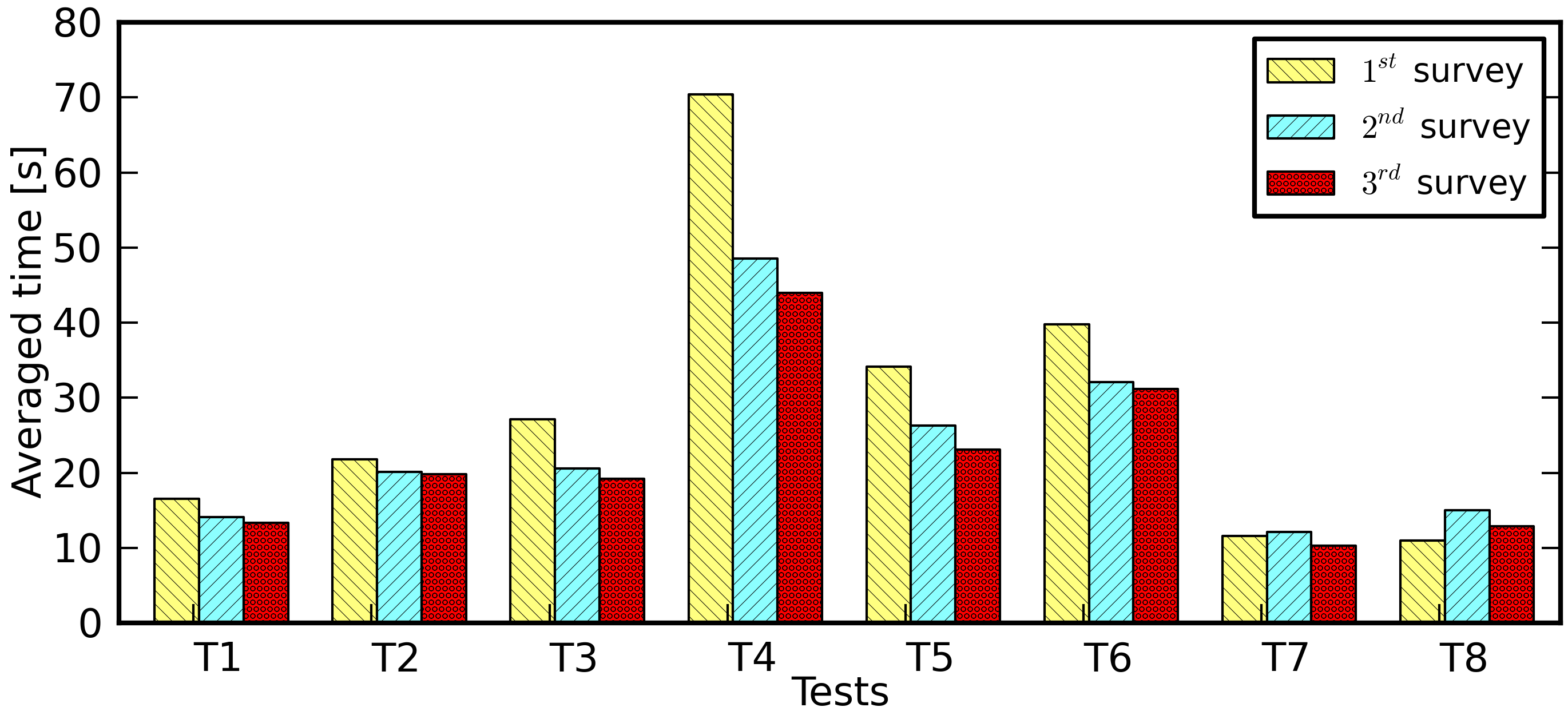}
\caption{Completion time.}
\label{fig:learning_effect_success}
\end{subfigure}
\caption{Learning effect on repeating the survey.}
\label{fig:learning_effect}
\end{figure}

Furthermore, we underline that in our user study, users were never trained before to solve our CAPTCHA, while trained users might need a smaller amount of time to solve CAPTCHaStar.
To verify this claim, we analyzed the performance of $25$ users that repeated the whole survey (i.e., tests from T1 to T8) at least three times.
The results of this analysis are reported in Figure~\ref{fig:learning_effect}. 
%

From Figure~\ref{fig:learning_effect}, we can observe that these users improved their performance to solve CAPTCHaStar challenges as they repeated the survey (i.e., increasing the success rate and decreasing of the completion time), while their performance on text-based challenges (T7 and T8) remained quite the same.
These preliminary results indicate that as users gain more confidence with CAPTCHaStar, the completion time significantly decreases.

\section{Resiliency to automated attacks}
\label{evaluateresiliency}
An important feature of a good CAPTCHA is the resiliency to automated attacks.
In the following, we investigate the resiliency of our proposal against several attacks, such as:
traditional attacks (Section~\ref{evaluatesecurity}); automated attacks using ad-hoc heuristics (Section~\ref{adhocattacks}) and attacks based on machine learning (Section~\ref{machinelearning}). 

\subsection{Traditional attacks}
\label{evaluatesecurity}
In this section, we discuss 
how CAPTCHaStar withstands traditional attack strategies for CAPTCHAs (we listed those strategies in Section~\ref{imagebasedthreats}).
\begin{itemize}
\item\textit{Indirect Attack} (A08): An indirect attack is not feasible, since all the information about the solution are not available on the client-side. 
CAPTCHaStar generates the challenge randomly on the server-side, and passes to the client only the description of the behavior of each star with respect to the current cursor position.
We remind that the coordinates $(sol_x, sol_y)$, corresponding to the 
solution of the challenge, are never revealed to the client.
Our system checks the correctness of the final answer on the server-side, only after the user confirms it. 

\item\textit{Exhaustion of Database} (A09): Our system generates a challenge starting from a \texttt{.png} picture, randomly chosen among more than five thousand candidates.
Moreover, this database can be automatically enriched with the help of a web crawler, but we consider this as a future work.

\item\textit{Leak of Database} (A10): An attacker who tries to match a challenge with its original picture faces a more complex problem than actually solving the challenge. 
Indeed, 
the attacker has to solve the challenge in order to input the complete shape to a matching algorithm.
Moreover, we highlight that during the generation phase the system alters the original picture, as described in Section~\ref{captchastaroverview}.

\item\textit{Machine Learning} (A11): In order to understand the feasibility of this attack, we actually trained a classifier to beat our CAPTCHA. Results suggest that this approach could be a serious threat, but it needs an unpractical amount of time and resources to be performed. We provide more detailed study about this specific attack in Section~\ref{machinelearning}.

\item\textit{Random Choice} (A12): For the sake of usability, CAPTCHaStar also accepts as a correct answer the neighborhood of the solution (according to the value of \textit{usability tolerance} parameter).
Nevertheless, the probability of success of a random guess is some $0.09$\% with \textit{usability tolerance} equal to $5$.
\item\textit{Pure Relay Attack} (A13): The solution discovery requires constant interaction with the CAPTCHA.
For this reason, a single screenshot sent to a third party is surely not enough to put a relay attack into practice.
\item\textit{Stream Relay Attack} (A14): As we introduced in Section~\ref{imagebasedthreats}, a stream relay attack needs to synchronously stream the current state to a human third party.
CAPTCHaStar needs a constant and immediate feedback system on each cursor movement.
Streaming a large number of frames over a (usually) slow connection between the bot and the solvers machine may reduce solving accuracy and increase the response time.
Unfortunately, this attack strategy remains the most effective against CAPTCHAs (including our proposal).
%
\end{itemize}

\subsection{Automated attacks using ad-hoc heuristics}
\label{adhocattacks}
In this section, we describe the design of a CAPTCHaStar automatic solver, in order to deeply test the reliability of our design. 
While retrieving all the possible states of a challenge is a trivial task (an attacker can simply take a snapshot for each cursor position), identifying the specific state corresponding to the solution is not simple. 
Indeed, the core task of an automatic solver is to recognize the presence of a shape in a given state. 
In the following, we report some ad-hoc heuristics we came up with to perform this task (of course, we cannot exclude better solutions that could be proposed in the future).

We created a program capable of generating 
every possible state, and assign a score to each state using a heuristic.
Given a state, the aim of the heuristic is to quantify the dispersion of the stars.
We consider as a candidate solution the state that minimizes the score given by the applied heuristic. 
The total number of states that the automatic solver has to evaluate is equal to $84100$ (i.e., $290^2$).
\hilight{In order to achieve the highest possible success rate, we chose to consider the whole research space, instead of a sub-sample.}
The computational cost of the attack can be really high, depending on the implemented heuristic.
We implemented the automatic solver and the heuristics described below using the \textit{C} programming language.
For each heuristic, we evaluate the automatic solver in terms of success rate and average execution time for at least $250$ challenges.
For this evaluation, we use the same value of parameters as in test T2 in the usability survey in Section~\ref{evaluateusability} 
(we chose these parameters since test T2 was the test with the highest success rate). 
In this evaluation, we used a PC with a $2.3$ GHz Intel Pentium B970 and $4$ GB memory.

\subsubsection{Minimize height/width of stars (\textit{MinSize})}
We define with $S^k$ the challenge state generated when the cursor is in position $k$, in coordinates $(x_k,y_k)$.
We also consider $x_s$ and $y_s$ the x and y coordinates, respectively, of star $s$.
The heuristic is defined as follows: 
$$ {MinSize}(k) = \left(\max_{s\in S^k } x_s - \min_{s\in S^k } x_s \right) + \left( \max_{s\in S^k } y_s- \min_{s\in S^k } y_s \right).$$

When $\psi = 0\%$, this heuristic has more than $90\%$ of success rate.
The addition of a few noisy stars to the challenge completely nullify the effectiveness of this heuristic (i.e., success rate of 0\% with only two noisy stars). 
The algorithm has a very low computational cost.
We recorded an average execution time of $10$ seconds.

\subsubsection{Minimize the distribution (\textit{MinDistribution})}
The main idea under this heuristic consists of dividing the drawable space into tiles.
Indeed, this heuristic evaluates the stars dispersion on each tile singularly. 
Henceforth, we define a matrix $M^k$ as the matrix of pixels in the drawable area, after the drawing process of the state $S^k$.
Each cell is defined as follows:
$$
M^k_{i,j}=
 \begin{cases}
      1 & \text{if pixel}\ (i,j) \text{ is white;}\  \\
      0 & \text{otherwise.}
    \end{cases} 
$$
We divide $M^k$ in a set $T^k$ of $144$ squared tiles (i.e., $t \in T^k$ is a sub-matrix of $M^k$), each with a side of 25 pixels.
We define the score of a single tile $t \in T^k$ as:
$$ f_{score}(t)=|2\cdot \sum_{i=1}^{25} \sum_{j=1}^{25} t_{ij}-25^2| .$$ 
The heuristic is defined as:
$$ {MinDistr}(k)= \sum_{t\in T^k} f_{score}(t) .$$
The value of the \textit{sensitivity} parameter ($\delta$) heavily affects the effectiveness of this heuristic.
Indeed, when $\psi=70\%$, the attack that uses this heuristic achieves a success rate of $2.7\%$, with $\delta=5$.
On the other hand, the success rate significantly decreases to $0.07\%$, when we increase the value of $\delta$ to $7$. 
The computational cost of this heuristic is slightly higher than the previously discussed \textit{MinSize}.
The average time is $65$ seconds. 

\subsubsection{Minimize the sum of distances (\textit{MinSumDist})}
This heuristic aims to detect when stars are clustered together, even in different groups. 
We define $d(s_1,s_2)$ as the euclidean distance between the stars $s_1$ and $s_2$.
The heuristic is defined as follows:
$$ {MinSumDist}(k)=\sum_{s\in S^k} \min_{r\in S^k}d(s,r) .$$
When $\psi=70\%$ and $\delta=7$, the success rate of this strategy is $0.56\%$.
The computational cost of this heuristic is higher than \textit{MinDistribution}:
we observed an average execution time of $12$ minutes and $45$ seconds.

\subsubsection{Minimize the sum of all distances (\textit{AllSumDist})}
We modify the previously discussed heuristic in order to consider all distances.
The heuristic is defined as:
$${AllSumDist}(k)=\sum_{s\in S^k} \sum_{r\in S^k} d(s,r).$$
%
This heuristic is the most effective, with a success rate of $1.92\%$ on $\psi=70\%$ and $\delta=7$.
However, this heuristic has a very high computational cost.
We recorded an average execution time of more than $25$ minutes. 
\hilight{Using sampling pairs on this algorithm would reduce the search space, and thus the execution time.
However, we underline that this attack will remain useless, since its performance is less than 2\% of success.
We expect that the use of sampling pairs would further reduce this success rate.}

\begin{figure}[ht]
\centering
\begin{subfigure}{.48\textwidth}
\centering
\includegraphics[scale=0.78]{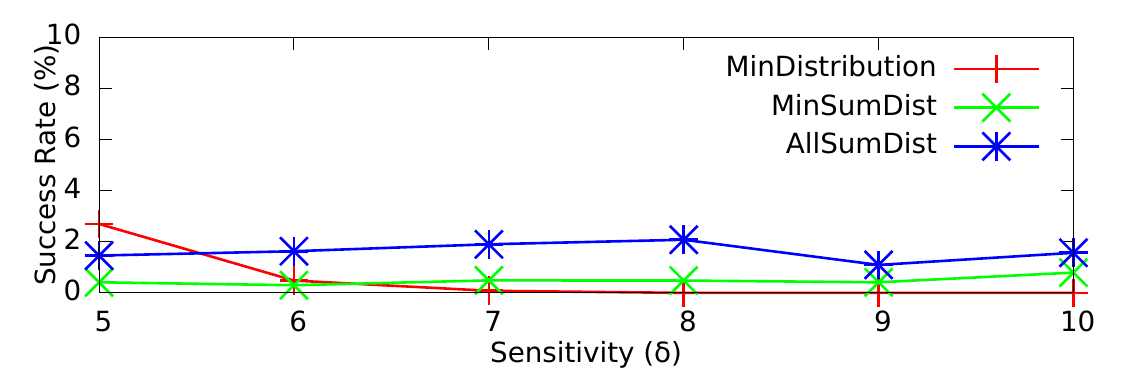}
  \caption{$\psi=70\%$,  varying $\delta$.} 
  \label{fig:allattacksONsens}
\end{subfigure}
 \begin{subfigure}{.48\textwidth}
\centering
\includegraphics[scale=0.78]{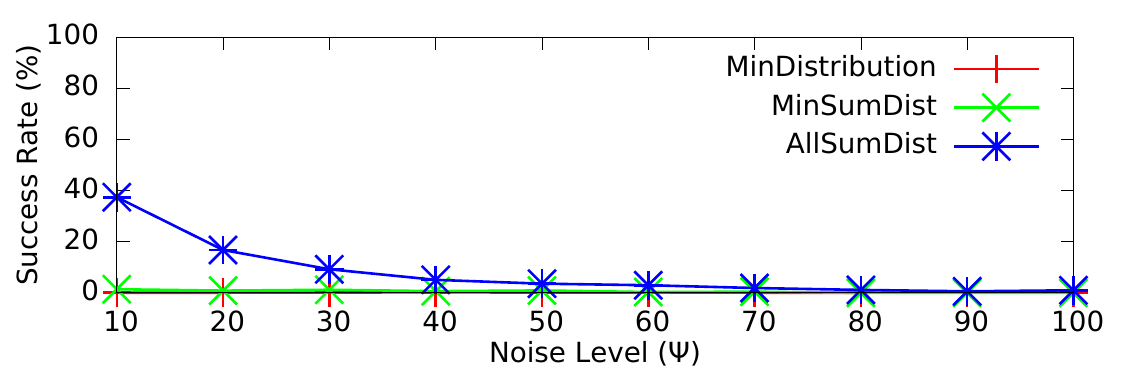}
 \caption{$\delta=7$, varying $\psi$.} 
\label{fig:allattacksONnoise}
\end{subfigure}
\caption{Comparison of success rates on variations of $\delta$ and~$\psi$.}
\label{fig:allattacksONnoiseANDsens}
\end{figure}

In Figure~\ref{fig:allattacksONsens} and Figure~\ref{fig:allattacksONnoise}, we report how the success rates of the heuristics described above vary at the change of parameters $\delta$ and $\psi$. 
From Figure~\ref{fig:allattacksONsens}, we observe that for $\psi=70\%$, the success rate is always smaller than $3\%$.
From Figure~\ref{fig:allattacksONnoise}, we observe that with a small level of noise the success rate would be significant (i.e., $40\%$ for $\psi=10$ and $\delta=7$).
However, increasing the noise level effectively mitigates this problem (i.e., for $\delta=7$ and $\psi>50\%$, the success rate is always smaller than $5\%$).

From Table~\ref{tab:exectimes}, we observe that even if the variation of execution time is very high (from $10$ seconds of MinSize, to $1500$ of AllSumDist ones), the success rate is always smaller than $2\%$.

\begin{table}[ht]
\centering
\scalebox{1.}{
\begin{tabular}{|>{\hspace{\tabred}}l<{\hspace{\tabred}}|
		>{\hspace{\tabred}}c<{\hspace{\tabred}}|
		>{\hspace{\tabred}}c<{\hspace{\tabred}}|
		>{\hspace{\tabred}}c<{\hspace{\tabred}}|
		>{\hspace{\tabred}}c<{\hspace{\tabred}}|} \hline
\textbf{Strategy} & \textbf{{MinSize}} & \textbf{{MinDistribution}} & \textbf{{MinSumDist}} & \textbf{{AllSumDist}} \\ \hline
\textbf{Time (s)} & $10$ & $65$ & $765$ & $1500$ \\ \hline
\textbf{Succ. Rate} & $0.00\%$ & $0.07\%$ & $0.50\%$ & $1.92\%$ \\ \hline
\end{tabular}
}
\caption{Execution time and Success ($\psi=70\%$; $\delta=7$).}
\label{tab:exectimes}
\end{table}

\subsection{Attacks based on machine learning} 
\label{machinelearning}
In order to assess the resiliency of CAPTCHaStar against machine learning-based attacks, we designed a tool that tries to find the solution of a challenge.
We implemented this tool using scikit-learn 
libraries~\cite{sklearn}.
In the following, we report in details how we built such tool. 
In particular, in Section~\ref{featuresextraction}, we introduce the methodology we followed to extract features from a challenge state;
in Section~\ref{classifiertraining}, we explain the training phase of the classifiers; 
and in Section~\ref{attackandperformance}, we describe the actual attack and we evaluate its performance. 

\subsubsection{Features extraction}
\label{featuresextraction}
We recall that given a state $S^k$, we obtain its Boolean matrix $M^k$, as defined in Section~\ref{adhocattacks}.
A classifier is a supervised learning algorithm~\cite{kotsiantis2007supervised} that requires a training set. 
The examples in the training set are labeled with the class they belong to. 
After the training phase, a classifier should be able to identify to which class a new unlabeled example belongs.
All the examples must have a fixed number of features.
Therefore, we need to represent a state of a challenge with a vector of $n$ features.
The methodology we follow for features extraction derives from
the procedure described in~\cite{golle2008machine}, but with a significant difference in the computation of features values. 
Indeed, we need to represent Boolean matrices (i.e., black and white) instead of gray-scale matrices. 
The idea consists of dividing a matrix $M^k$ into a set $T^k_\omega$ of squared tiles. 
The parameter $\omega$ is the amount of pixels in a tile side.
For each considered value of $\omega$, we build a vector $F_\omega=<f_1,..,f_n>$ of reference tiles.
In particular, we empirically select $n = 3 \omega$. 
From now on, we refer to $h(t_1,t_2)$ as the Hamming distance between two Boolean matrices $t_1$ and $t_2$ (i.e., two tiles).
The tiles in the vector $F_\omega$ must be different from each other. 
For this reason, 
we apply k-mean clustering method on a training set of candidate tiles with side $\omega$, using $K=n$ and $h$ as similarity metric.
At the end of the clustering procedure, we obtain a vector $F_\omega$, where $f_i \in F_\omega$ is a tile that represents the centroid of the $i^{th}$ cluster.
We compute the values of a vector $D^k_\omega=<d^k_1,..,d^k_n>$, where each value is defined as follows: 
\newcommand{\argmin}{\operatornamewithlimits{argmin}}
\newcommand{\argmax}{\operatornamewithlimits{argmax}}
$$d^k_{i} = | \{ t \in T^k_{\omega} : f_i = \argmin_{{l}\in F_\omega}\ h({l},t) \}|,\ \ \forall i = 1,..,n.$$ 
The values in the features vectors $D^k_\omega$ are normalized, from $0$ to $1$. 
In practical terms, for a fixed $\omega$, this procedure produces a vector of features $D^z_\omega$, starting from a cursor position $z$ that corresponds to the state $S^z$. 

\subsubsection{Classifiers training}
\label{classifiertraining}
We trained Random Forest (RF) and Support Vector Machine (SVM) classifiers with $4000$ random challenges (with $\psi=70\%$ and $\delta=7$).
For implementation of the classifiers, we use a RF classifier with $60$ Decision trees estimators, and we use an SVM classifier with Radial Basis Function (RBF) as the kernel function.
We use these classifiers to perform a binary classification, i.e., they recognize examples of two classes: \textit{solution} and \textit{non-solution}.
For each challenge, we generate $400$ states (this means a training set of $1.6 \cdot 10^6$ examples).
We train a classifier for each value of $\omega$.

We underline that an attacker has to build this training set manually, i.e., we have access to the exact solution of a challenge, while an adversary can retrieve this information only by solving the challenge legitimately.
Moreover, we also know the value of \textit{usability tolerance} parameter, which allows us to provide the neighbors of the solution to the classifier.

\subsubsection{Attacks design and performance}
\label{attackandperformance}
\begin{figure}[ht!]
\centering
\begin{subfigure}{.45\textwidth}
\includegraphics[width=0.99\textwidth]{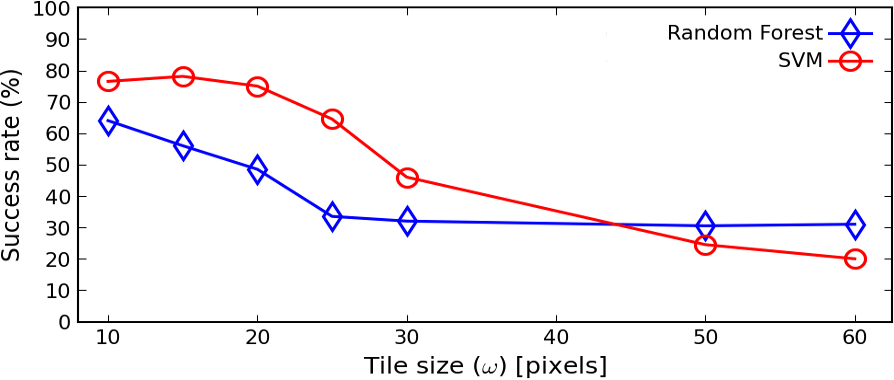}
\caption{Success rate.}
\label{fig:successRateML}
\end{subfigure}
\\
\begin{subfigure}{.45\textwidth}
\includegraphics[width=0.99\textwidth]{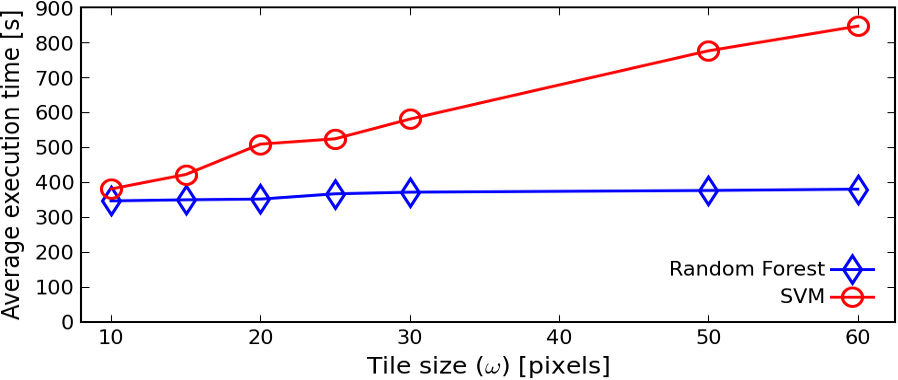}
\caption{Average execution time.}
\label{fig:timeML}
\end{subfigure}
\vspace{-0.cm}
\caption{Success rate and average execution time for machine learning-based attacks, varying the tile size ($\omega$).}
\label{fig:MLSuccess}
\end{figure}

In the following, we discuss the design of two attacks that use the classifiers trained in the previous phase.
We evaluated the attacks as the value of parameter $\omega$ changes.
For the sake of attack feasibility (in terms of both memory and time), we limited the research space to a subset of $K$ possible cursor positions coordinates, defined as: 
$$K_{\lambda}=\{ (\lambda x, \lambda y) : \forall x,y \in \mathbb{N} \cap [0,{300}/{\lambda}]\}.$$
We set the parameter $\lambda=5$ pixels (i.e., the same value as the \textit{usability tolerance}), in order to ensure that we have at least one solution among all the states.
After this procedure, we obtain a set of $K_{\lambda}$ cursor positions. 
For each classifier with a specific $\omega$, we define $C_\omega$ as the function that evaluates the probability that a given state $S^k$ belongs to the class \textit{solution}. 
We recall that a challenge admits only one answer, and it is final and irrevocable.
We observed experimentally that the distribution of values for function $C_\omega$ 
often presents multiple local maximums and large plateaus. 
For this reason, an attacker must find the cursor position $k_{sol}$ that corresponds to a global maximum for the function $C_\omega$: 
$$ k_{sol}=\argmax_{k \in K_{\lambda}} \ C_\omega(S^k). $$
In this evaluation, we ran the attacks on a test set of $200$ challenges (with $\psi=70\%$ and $\delta=7$), for each considered value of $\omega$. 
We executed the attacks on a high end PC with a $3.16$ GHz Intel Xeon X5460 and $32$ GB of RAM.
Figure~\ref{fig:successRateML} and Figure~\ref{fig:timeML} report the success rate and the average execution time to perform these attacks, respectively.
The attack with the best success rate uses the SVM classifier with $\omega=15$, and it achieves a success rate of $78.1\%$ (as reported in Figure~\ref{fig:successRateML}).

The time required to build the features vectors $D_{\omega}^{k},\ \forall k\in K_{\lambda}$, remains stable at around $340$ seconds.  
On one hand, the time required to compute the probability values $C_{\omega}(S^{k}),\ \forall k\in K_{\lambda}$, increases linearly using SVM classifier, according to the value of $\omega$.
On the other hand, this time remains under two seconds using RF. 
%
This means that an attack on a single challenge will have some $78\%$ of success rate, but it will require $421$ seconds to be performed.
We recall that a human user can solve a challenge with a success rate of more than $90\%$ in an average time of $27$ seconds ($56$ seconds in the worst case).
Therefore, the problem for a bot of automatically recognizing a solution state of a challenge of CAPTCHaStar is hard to treat in a limited amount of time and resources.
We underline that, as recently reported in~\cite{bursztein2014end}, machine learning based attacks achieve some $50\%$ success rate in only two seconds against Baidu and eBay CAPTCHAs.
\newcommand*\rot{\rotatebox{90}}
\newcommand*\OK{\ding{51}}
\newcommand*\NO{\ding{55}}

\section{Discussion}
\label{discussion}
In this section, we first compare our solution with other image-based CAPTCHAs (Section~\ref{comparison}), then we discuss limitations of CAPTCHaStar (Section~\ref{limitations}) and finally we discuss some future work (Section~\ref{futurework}). 

\subsection{Comparison with other image-based CAPTCHAs }
\label{comparison}
Comparing our solution with the state-of-the-art of image-based CAPTCHAs (presented in Section~\ref{imagebasedcompetitors}), our proposal turns out to be more resilient against attacks. 
In particular, Table~\ref{tab:competitors} reports the comparison considering the common weaknesses of image-based CAPTCHAs, previously discussed in Section~\ref{imagebasedthreats}.
In the table we indicate whether the design is protected against the following attacks: \textit{indirect attack}, \textit{exhaustion of DB}, \textit{leak of DB}, and \textit{pure relay attack}.
In addition, for \textit{stream relay attack} and \textit{machine learning} based attacks, we report the cost to perform such attack, in terms of computational time and resources.

We notice that most of the designs in the literature limit their focus to a specific threat, but they offer less protection against others.
On the other hand, our proposal is designed to resist all of them, while maintaining a high usability level. 
%
\begin{table}[h!]
\newcommand{\specialcell}[2][c]{%
  \begin{tabular}[#1]{@{}c@{}}#2\end{tabular}}
\centering
\scalebox{0.94}{
\begin{tabular}{ | c | c@{\hskip 0.5mm}c | c@{\hskip 0.5mm}c | c@{\hskip 0.5mm}c | c@{\hskip 0.5mm}c | c@{\hskip 0.5mm}c | c@{\hskip 0.5mm}c | r@{\hskip 0.5mm}l |}
\hline
    \specialcell[b]{\textbf{CAPTCHA}\\\textbf{design}} & \rot{\textbf{Indirect}} & \rot{\textbf{attack}} & \rot{\textbf{Exhaustion}} & \rot{\textbf{of the DB}} & \rot{\textbf{Leak}} & \rot{\textbf{of the DB}} &  \rot{\textbf{Pure relay}} & \rot{\textbf{attack}} & \rot{\textbf{Stream relay}} & \rot{\textbf{attack}} & \rot{\textbf{Machine}} & \rot{\textbf{learning}} & \rot{\textbf{Random}} & \rot{\textbf{chance}} \\ \hline
	Asirra~\cite{elson2007asirra} & \multicolumn{2}{c|}{\OK} & \multicolumn{2}{c|}{\OK} & \multicolumn{2}{c|}{\NO} & \multicolumn{2}{c|}{\NO} & \multicolumn{2}{c|}{low} & \multicolumn{2}{c|}{low} & 0{\hskip -0.5mm} & .02\%  \\
	Collage~\cite{shirali2008advanced} & \multicolumn{2}{c|}{\OK} & \multicolumn{2}{c|}{\NO} & \multicolumn{2}{c|}{\NO} & \multicolumn{2}{c|}{\NO} & \multicolumn{2}{c|}{low} & \multicolumn{2}{c|}{high} & 16{\hskip -0.5mm} & .60\% \\
	Deep~\cite{nejati2014deepCAPTCHA}    & \multicolumn{2}{c|}{\OK} & \multicolumn{2}{c|}{\OK} & \multicolumn{2}{c|}{\OK} & \multicolumn{2}{c|}{\NO} & \multicolumn{2}{c|}{low} & \multicolumn{2}{c|}{high} & 0{\hskip -0.5mm} & .20\%  \\
	Motion~\cite{shirali2008motion}  & \multicolumn{2}{c|}{\OK} & \multicolumn{2}{c|}{\NO} & \multicolumn{2}{c|}{\NO} & \multicolumn{2}{c|}{\OK} & \multicolumn{2}{c|}{low} & \multicolumn{2}{c|}{high} & 25{\hskip -0.5mm} & .00\%  \\
	Video~\cite{kluever2009balancing} 	& \multicolumn{2}{c|}{\OK} & \multicolumn{2}{c|}{\OK} & \multicolumn{2}{c|}{\NO} & \multicolumn{2}{c|}{\NO} & \multicolumn{2}{c|}{low} & \multicolumn{2}{c|}{high} & 0{\hskip -0.5mm} & .30\% \\
	Noise~\cite{okada2012new} 	& \multicolumn{2}{c|}{\OK} & \multicolumn{2}{c|}{\OK} & \multicolumn{2}{c|}{\OK} & \multicolumn{2}{c|}{\OK} & \multicolumn{2}{c|}{mid} & \multicolumn{2}{c|}{mid} &$\sim 0${\hskip -0.5mm} & .00\% \\
	Cursor~\cite{thomas2013cursor}  & \multicolumn{2}{c|}{\OK} & \multicolumn{2}{c|}{\OK} & \multicolumn{2}{c|}{\OK} & \multicolumn{2}{c|}{\OK} & \multicolumn{2}{c|}{low} & \multicolumn{2}{c|}{low} & $\sim 0${\hskip -0.5mm} & .00\% \\
	Jigsaw~\cite{gao2010novel}  & \multicolumn{2}{c|}{\OK} & \multicolumn{2}{c|}{\NO} & \multicolumn{2}{c|}{\NO} & \multicolumn{2}{c|}{\NO} & \multicolumn{2}{c|}{low} & \multicolumn{2}{c|}{mid} & 6{\hskip -0.5mm} & .66\% \\
	PlayThru~\cite{areyouahuman} & \multicolumn{2}{c|}{\NO} & \multicolumn{2}{c|}{\NO} & \multicolumn{2}{c|}{\NO} & \multicolumn{2}{c|}{\OK} & \multicolumn{2}{c|}{high} & \multicolumn{2}{c|}{high} &$\sim 0${\hskip -0.5mm} & .00\% \\
	\textbf{CAPTCHaStar}   	& \multicolumn{2}{c|}{\textbf{\OK}} & \multicolumn{2}{c|}{\textbf{\OK}} & \multicolumn{2}{c|}{\textbf{\OK}} & \multicolumn{2}{c|}{\textbf{\OK}} & \multicolumn{2}{c|}{\textbf{high}} & \multicolumn{2}{c|}{\textbf{high}} & \textbf{ 0}{\hskip -0.5mm} & \textbf{.09\%} \\
    \hline
\end{tabular}
}
\caption{Protection against the threats in Section~\ref{imagebasedthreats}.}
\label{tab:competitors}
\end{table}

\subsection{Limitations}
\label{limitations}
Unfortunately, we (as well as other CAPTCHA proposers) are not able to “prove” that our CAPTCHA is secure against all the possible attacks. We believe the best a researcher can do in such cases is to consider both current traditional attacks (see our Section~\ref{evaluatesecurity}) and ad-hoc ones (see our sections~\ref{adhocattacks} and~\ref{machinelearning}). 
As an example, Asirra~\cite{elson2007asirra} 
has been later proven to be breakable~\cite{golle2008machine,zhu2010attacks}. 
The same goes for other CAPTCHAs, such as ReCaptcha (both versions of 2011 and 2013) and the ones employed by CNN, Wikipedia, Yahoo, Microsoft~\cite{yan2008low} and PlayThru~\cite{areyouahuman}.

In this paper, we evaluated the attacks that we were able to think up with our best efforts. 
Such evaluation suggests that our proposal is still more resilient than some state-of-the-art CAPTCHAs.
Indeed, BAIDU and eBay CAPTCHAs can be broken with a success rate of 50\% in just two seconds~\cite{bursztein2010good}). 

We demonstrated that the problem of classifying an arrangement of dots as forming a shape or not is learnable by a SVM classifier.
However, as shown by the attack evaluation reported in Section~\ref{machinelearning}:
the bottleneck for an attack, that uses such shape recognition as a building block, is the generation of the sampled search space (i.e., the possible configurations by varying the mouse coordinates).
Finally, such complexity can be further increased by enlarging the drawable space (also reducing the success rate of random chance attack).

%
\subsection{Future work}
\label{futurework}
As a future work, we plan to further increase the resiliency of CAPTCHaStar by analyzing the pattern of mouse movements during the resolution of a challenge.
We believe this analysis will be meaningful in order to better discriminate human users and automatic programs.

Moreover, we intend to perform some qualitative experiments in the laboratory, especially to evaluate the usability on mobile devices.
\hilight{However, we highlight that CAPTCHaStar already takes into account the recommendations to improve CAPTCHA usability on smartphones, recently proposed by Chiasson et al.} in~\cite{chiasson15}.  
In fact, our proposal has the following properties, which are desirable features for CAPTCHAs on a smartphone:
\begin{itemize}[noitemsep,topsep=0pt]
 \item Our design focuses only on a single task, avoiding optional features. 
 \item The input mechanism of CAPTCHaStar on touch-enabled devices (described in Section~\ref{user_interaction}) is cross-platform, and it does not interfere with normal operations of the browser.
 \item It is possible to isolate CAPTCHaStar from the rest of the web form.
 \item CAPTCHaStar minimizes bandwidth usage. Indeed, the data transmission with the server is limited to the exchange of trajectories parameters of the stars, and the verification of the user final answer.
 \item Our design does not require any additional feature nor library to be installed on the browser.
\end{itemize}

Finally, we plan to investigate the possibility of leveraging additional gaps between human abilities and automatic programs. 
For example, we intend to involve in a challenge the semantic meaning of the final shape.
This means to rely on the innate human ability to relate objects with their semantic. 
In fact, nowadays this ability is hardly imitable by a machine~\cite{zhu2010attacks}.
We strongly believe that improving CAPTCHaStar challenges in this way will further increase the resiliency of our proposal against machine learning-based attacks.

\section{Conclusions}
\label{conclusions}
In this paper, we proposed CAPTCHaStar, a novel image-based CAPTCHA that leverages the innate human ability to recognize shapes in a confused environment~\cite{kanizsa1979organization}.
Our study demonstrates that our proposal meets both security and usability requirements for a good 
CAPTCHA design.

We described in detail our prototype implementation and the selection of parameters involved in challenge generation. 
Data collected from our user study confirmed the usability of our proposal.
Indeed, users were able to obtain a success rate higher than $90\%$, which is better than the success rates of CAPTCHAs currently used in websites~\cite{bursztein2010good} such as mail.ru and Microsoft.
Finally, the majority of the users who participated in our survey preferred CAPTCHaStar over classical text-based CAPTCHAs.
These results motivate further research in this direction.

In this paper, we also assessed the resiliency of CAPTCHaStar against traditional and automated ad-hoc attacks.
Indeed, these attacks were shown to be ineffective (for a proper setting of our parameters). 
We also performed an attack leveraging a machine learning classifier, which we optimized by reducing as much as possible the research space.
Despite this optimization of the attack and its execution on a high end PC, the resulting average execution time is still unacceptable,
i.e., more than six minutes to find the solution for a single challenge (with a success probability of $79\%$).
We recall that users are able to complete CAPTCHaStar challenges in an average time of less than $27$ seconds (with a success probability of some $90\%$).
Furthermore we recall that attacks to the state of the art CAPTCHAs take only two seconds~\cite{bursztein2014end} (with a probability of some $50\%$).
%
%
%
%
\balance
\bibliographystyle{abbrv}

%
%
%
\end{document}